\documentclass{elsart}
\usepackage{graphicx,harvard,amssymb}

\makeatletter
\def\elsartstyle{%
	\def\normalsize{\@setfontsize\normalsize\@xiipt{14.5}}
	\def\small{\@setfontsize\small\@xipt{13.6}}
	\let\footnotesize=\small
	\def\large{\@setfontsize\large\@xivpt{18}}
	\def\Large{\@setfontsize\Large\@xviipt{22}}
	\skip\@mpfootins = 18\p@ \@plus 2\p@
	\normalsize
}
\makeatother
\def\bibcode#1{(\texttt{#1})}

\def\url#1{{\ttfamily\def\/{/\discretionary{}{}{}}#1}}

\begin{document}
\begin{frontmatter}
\title{Cluster Lensing of the CMB}
\author[address1]{Chris Vale\thanksref{cvemail}}, 
\author[address2]{Alexandre Amblard\thanksref{aaemail}}, 
\author[address1,address2]{Martin White\thanksref{mwemail}}
\address[address1]{Department of Physics, University of California,
Berkeley, CA, 94720}
\address[address2]{Department of Astronomy, University of California,
Berkeley, CA, 94720}
\thanks[cvemail]{E-mail: cvale@astro.berkeley.edu}
\thanks[aaemail]{E-mail: amblard@astro.berkeley.edu}
\thanks[mwemail]{E-mail: mwhite@astro.berkeley.edu}
\begin{abstract}
We investigate what the lensing information contained in high
resolution, low noise CMB temperature maps can teach us about cluster
mass profiles.  We create lensing fields and Sunyaev-Zel'dovich effect
maps from N-body simulations and apply them to primary CMB
anisotropies modeled as a Gaussian random field.  We examine the
success of several techniques of cluster mass reconstruction using CMB
lensing information, and make an estimate of the observational
requirements necessary to achieve a satisfactory result.
\end{abstract}
\begin{keyword}
Cosmology: Cosmic Microwave Background, Cosmology: Theory,
Cosmology: Gravitational Lensing, Cosmology: Large-Scale Structure of Universe
\PACS 98.80$-$k
\end{keyword}
\end{frontmatter}

\section {Introduction} \label {sec:intro}

The study of anisotropies in the Cosmic Microwave Background (CMB) has
proven to be a gold-mine for cosmology.  The primary anisotropies on scales
larger than $10'$ have now been probed with high fidelity by WMAP \cite{WMAP}
over the whole sky, leading to strong constraints on our cosmological model.
Within the next few years this activity will be complemented by high
angular resolution, high sensitivity observations of secondary anisotropies
by the SZA\footnote{http://astro.uchicago.edu/sza/},
APEX-SZ experiment\footnote{http://bolo.berkeley.edu/apexsz/},
the South Pole Telescope (SPT\footnote{http://astro.uchicago.edu/spt/})
and the Atacama Cosmology Telescope
(ACT\footnote{http://www.hep.upenn.edu/$\sim$angelica/act/act.html})
which are aiming to make arcminute resolution maps with $10 \mu$K sensitivity
(or better) at millimeter wavelengths.

On the angular scales the dominant secondary anisotropy is expected to be
the Compton scattering of cold CMB photons from hot gas along the line of
sight, known as the thermal Sunyaev-Zel'dovich (SZ) effect
(\cite{SZ72,SZ80}; for recent reviews see
\citeasnoun{Rep} and \citeasnoun{Bir}).
The thermal SZ effect can be spectrally distinguished from primary CMB
anisotropies given enough sensitivity and frequency coverage, and we shall
not consider it in this work.
At slightly lower amplitudes are the kinetic SZ effect and gravitational
lensing, both of which leave the CMB spectrum unaltered but modify the
spatial correlations and statistics of the signal.
These effects can in principle provide useful constraints on re-ionization
models (the kSZ effect, e.g.~\cite{zhang03})
and allow us to map the dark matter back to the surface of last scattering
(lensing, e.g.~\cite{seljak99,zaldarriaga99,hu01,hirata03,okamoto03}).

In this paper we want to consider gravitational lensing of the CMB by
galaxy clusters.
This was first studied by \citeasnoun{SZ00},
which is the starting point for our work.  Relatively little other work
has been done on this phenomenon, notable exceptions being the work of
\citeasnoun{Coo03} who described a method to measure the equation of state
of the dark energy, \citeasnoun{Bar03} who gave CMB lensing as an
example of numerical techniques and \citeasnoun{HK04} whose
aim was quite similar to the work presented here.

Our goal is to study how well, and in what manner, we can reconstruct the
cluster profile, or an integrated quantity such as the mass, from the
lensing induced distortion in the CMB temperature field\footnote{We shall
only consider lensing of the temperature anisotropies in this work,
neglecting polarization.  This is partly motivated by the fact that some
of the upcoming experiments will not be polarization sensitive, and partly
to keep the calculation under control.},
or conversely to understand the impact of large collapsed structures on
the statistics of the CMB.
The principle advantage of CMB lensing over traditional lensing of galaxies
is that the source redshift is almost perfectly known.  The main
disadvantage is that it is presents a single, fixed source plane.
Lensing also represents auxiliary science that can be done with already
planned or funded instruments, at little or no additional cost.  As such
it is worth investigating in some detail.

It has become well known that lensing suffers from severe projection effects
\cite{RebBar,MetWhiLok}
so we shall here consider how well the {\it projected\/} mass profiles can
be constructed from CMB lensing, leaving aside the question of how well such
profiles can be deprojected to get 3D quantities.
In order of decreasing desirability we would like to reconstruct the
convergence map (projected density) of every cluster in the field;
compute the total convergence (mass) of every cluster in the field;
compute the profile of `stacked' clusters or compute some integral
of the `stacked' profile.
We shall investigate each of these in turn.

The plan of the paper is as follows.
In \S\ref{sec:theory} we introduce the lensing formalism, largely following
\citeasnoun{SZ00}, and introduce our notation.
We implement the S\&Z programme in \S\ref{sec:toy}, where we show how the
procedure works in a simple toy model of an ideal cluster lensing a pure,
known CMB gradient.
Then we begin to add complications in \S\ref{sec:beyond},
looking in particular at the fact that the CMB is not a pure gradient,
the contamination from kSZ
(which is highly correlated with the lensing structures) and the
non-Gaussianity of the lensing field and finally at the effects of noise.
We summarize with our conclusions in \S\ref{sec:conclusion}.
Some details of the simulations we use to make mock observations are
given in an Appendix.

\section {The Theory of Cluster Lensing of the CMB} \label {sec:theory}

In this section we review the effect of cluster lensing on the CMB.
Our goal is to explore to what extent the mass and the mass profile of a
cluster may be constrained using information from high resolution
temperature maps of the CMB if the cluster's redshift and position on the
sky are known.  We introduce the formalism of CMB lensing but provide only
a brief summary of equations directly relevant here;
see \citeasnoun{BS01}
for a comprehensive review of weak lensing.
Throughout this paper we work in the weak lensing limit, assume a
flat $\Lambda \rm CDM$ universe, adopt units where where the speed of
light $c=1$, and work in comoving coordinates.  

\subsection{Weak Lensing of the CMB} \label{sec:formalism}

We begin by examining the gravitational lensing of light rays that
originate at the surface of last scattering by inhomogeneities in the
intervening matter distribution.  We define the primordial CMB
temperature field at the surface of last scattering as $\tilde
T$({\boldmath $\theta $}$'$) at an angular position on the sky
{\boldmath $\theta $}$'$.  Lensing by large scale structures such as
clusters will cause CMB light rays that originate at a position
{\boldmath $\theta $}$'$ to be deflected by an angle
$\delta${\boldmath $\theta $} to an observed position on the sky
{\boldmath $\theta $}, so that the observed temperature $T$({\boldmath
$\theta$}) is

\begin{equation}
\label{eq:cmb1}
  T(\mbox{\boldmath $\theta$}) = \tilde{T}(\mbox{\boldmath $\theta$}')
  = \tilde{T}(\mbox{\boldmath $\theta$} - \delta\mbox{\boldmath $\theta$})
\end{equation}

It is simple to derive a mathematical expression for the deflection
angle $\delta${\boldmath $\theta$} in the weak lensing limit.  The
total deflection angle $\delta${\boldmath $\theta$} of a source at
position $\chi_s$ as seen by an observer at $\chi = 0$ is

\begin{equation}
\label{eq:dtheta}
  \delta \mbox{\boldmath $\theta$} =
  {2 \over \chi_s} \int_0^{\chi_s} d\chi \ (\chi_s-\chi)\nabla_{\perp}\phi
\end{equation}

where $\nabla_\perp$ denotes the spatial gradient perpendicular to the
path of the light ray, $\phi$ is the three-dimensional peculiar
gravitational potential, and $\chi$ is the radial comoving coordinate.

As we discuss below, we will be interested in the dependence of the
lensing angle on the matter distribution.  To see this dependence, we
define the convergence $\kappa \equiv {1 \over 2}\nabla_\theta \cdot
\delta${\boldmath $\theta$} where $\nabla_\theta$ is the angular
gradient operator.  Then

\begin{equation} \label{eq:kapphi}
  \kappa \approx \int_0^{\chi_s} d \chi \ g(\chi, \chi_s) \nabla^2 \phi
\end{equation}

where we have invoked the Born and Limber approximations
(see \citeasnoun{JSW} and \citeasnoun{ValWhi} for a discussion)
and made use of the lensing kernel

\begin{equation} \label{eq:kernel}
  g(\chi, \chi_s) \equiv {(\chi_s - \chi )\chi \over \chi_s} 
\end{equation}

The matter density $\rho$ is related to the three-dimensional gravitational
potential $\phi $ through 

\begin{equation} \label{eq:poisson2}
  \nabla^2 \phi = 4 \pi G \bar \rho_0 {\delta \over a}
\end{equation}

where all quantities are defined with respect to comoving coordinates,
$G$ is the gravitational constant, $\bar \rho_0$ is the mean density
of the present day universe, and $\delta \equiv \rho / \bar \rho_0 -1$
is the relative mass overdensity.  It is worth noting that if only
local lensing information is available, equation (\ref{eq:poisson2})
will be uncertain up to an overall constant.  This is the source of
the so called mass sheet degeneracy.  We are now in a position to
relate $\kappa$ to the matter density by combining equations
(\ref{eq:kapphi}) and (\ref{eq:poisson2})

\begin{equation}
\label{eq:kapdelta}
\kappa( \mbox{\boldmath $\theta$} ) \approx 
4 \pi G \bar \rho_0 \int_0^{\chi_s} 
d \chi \ g(\chi, \chi_s) \ {\delta(\chi, \mbox{\boldmath $\theta$} ) \over a}
\end{equation}

If the lensing effect is primarily due to a single structure whose
size is much less than its comoving distance $\chi$, one can make use
of the thin lens approximation.  Then the convergence is simply
related to the projected two dimensional mass density
$\Sigma \equiv \int \rho d \chi$ of the lens, so that
equation (\ref{eq:kapdelta}) becomes

\begin{equation}
\label{eq:thin}
  \kappa(\mbox{\boldmath $\theta$}) \approx 
  4 \pi G  \bar \rho_0 g(\chi,\chi_s) \Sigma(\chi, \mbox{\boldmath $\theta$})
\end{equation}

We make use of equation~(\ref{eq:thin}) to create the convergence
field from the N-body simulations
(e.g.~Figure \ref{fig:clusters}), as detailed in the Appendix.
{}From the convergence field we compute the deflection angle via
\begin{equation}
  \delta\mbox{\boldmath $\theta$} = \nabla_\perp\nabla^{-2}\kappa
\end{equation}
and hence the lensed temperature field.

\begin{figure} 
\leftmargin=2pc
\begin{center}
\includegraphics*[height=6cm]{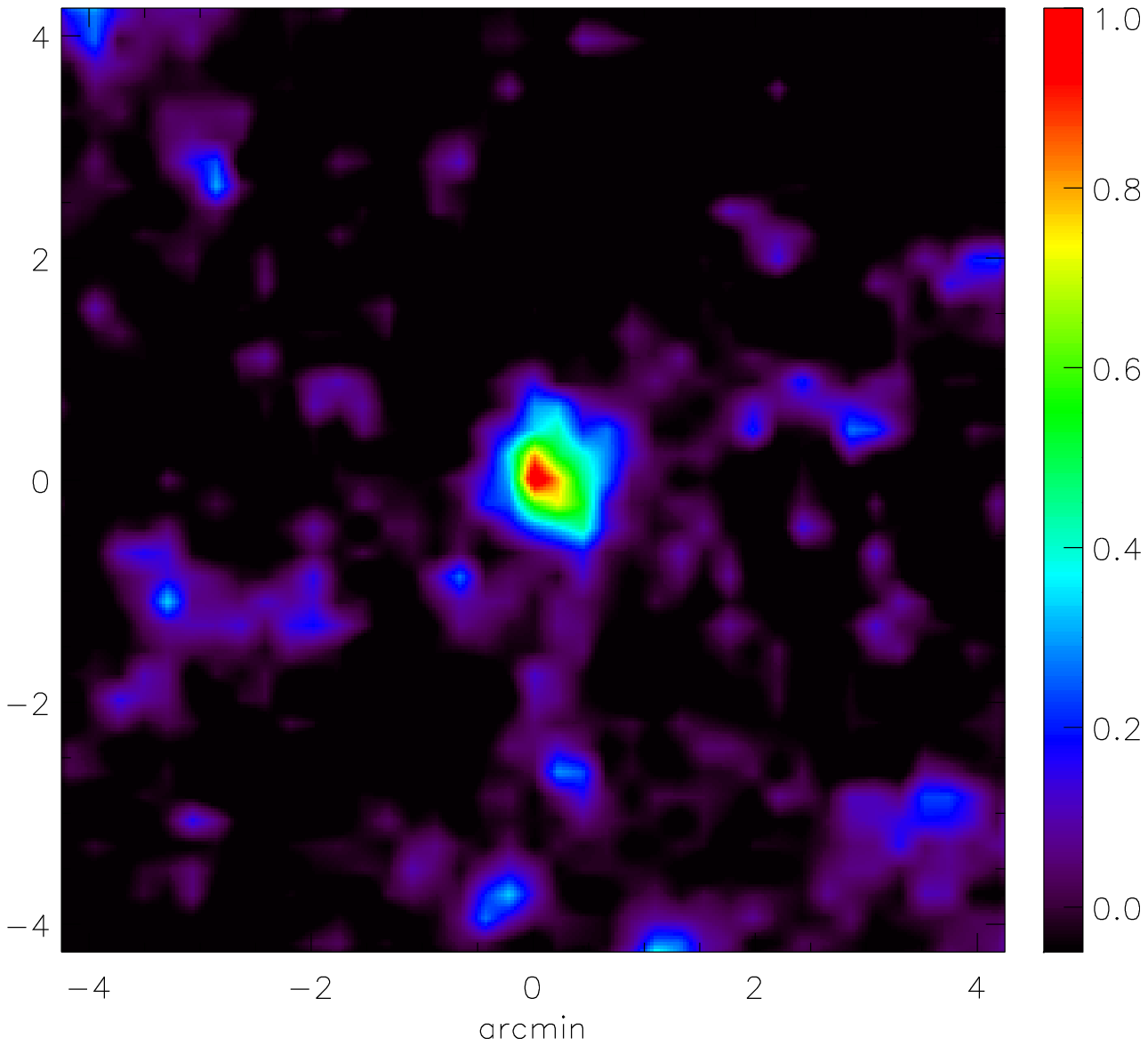}
\includegraphics*[height=6cm]{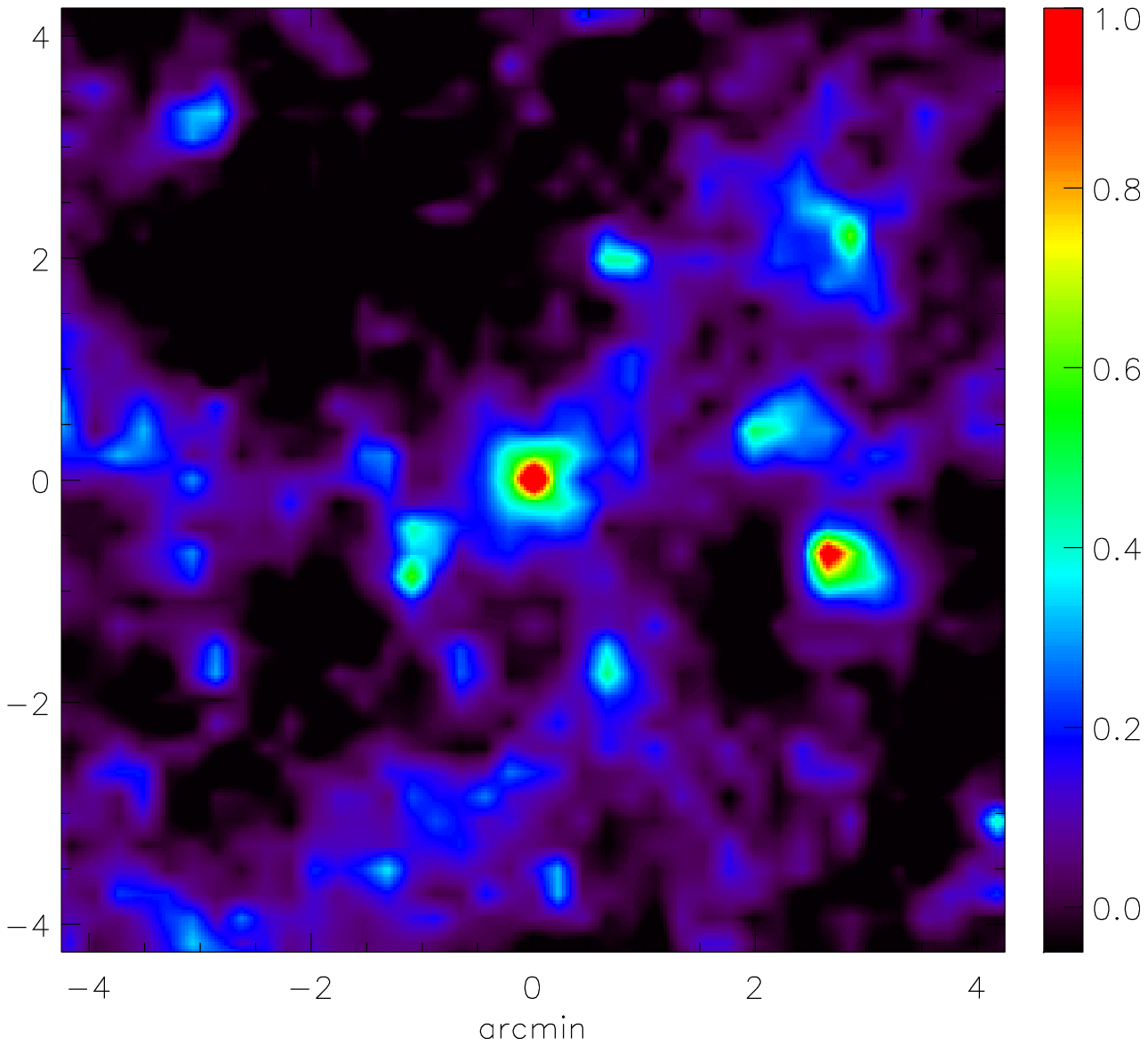}
\end{center}
\caption{The convergence maps for isolated and non-isolated clusters
as found in our simulations.  The color bar shows the value of the
(dimensionless) convergence $\kappa$.}
\label{fig:clusters}
\end{figure}

\subsection{Lensing by an Ideal Cluster} \label{sec:lensing}

It is instructive to examine the lensing effect of an isolated cluster
in the absence of other secondary anisotropies, foregrounds,
instrument effects, or lensing by other structures, all of which we
include later.  If the deflections are small, we may expand the right
hand side of equation~(\ref {eq:cmb1}) to linear order, so that

\begin{equation}
\label{eq:cmb2}
T(\mbox{\boldmath $\theta$})
  \approx \tilde{T}(\mbox{\boldmath $\theta$})
  - \delta\mbox{\boldmath $\theta$}
    \cdot\nabla\tilde{T}(\mbox{\boldmath $\theta$} )
\end{equation}

\begin{figure} 
\leftmargin=2pc
\begin{center}
\includegraphics*[width=6cm]{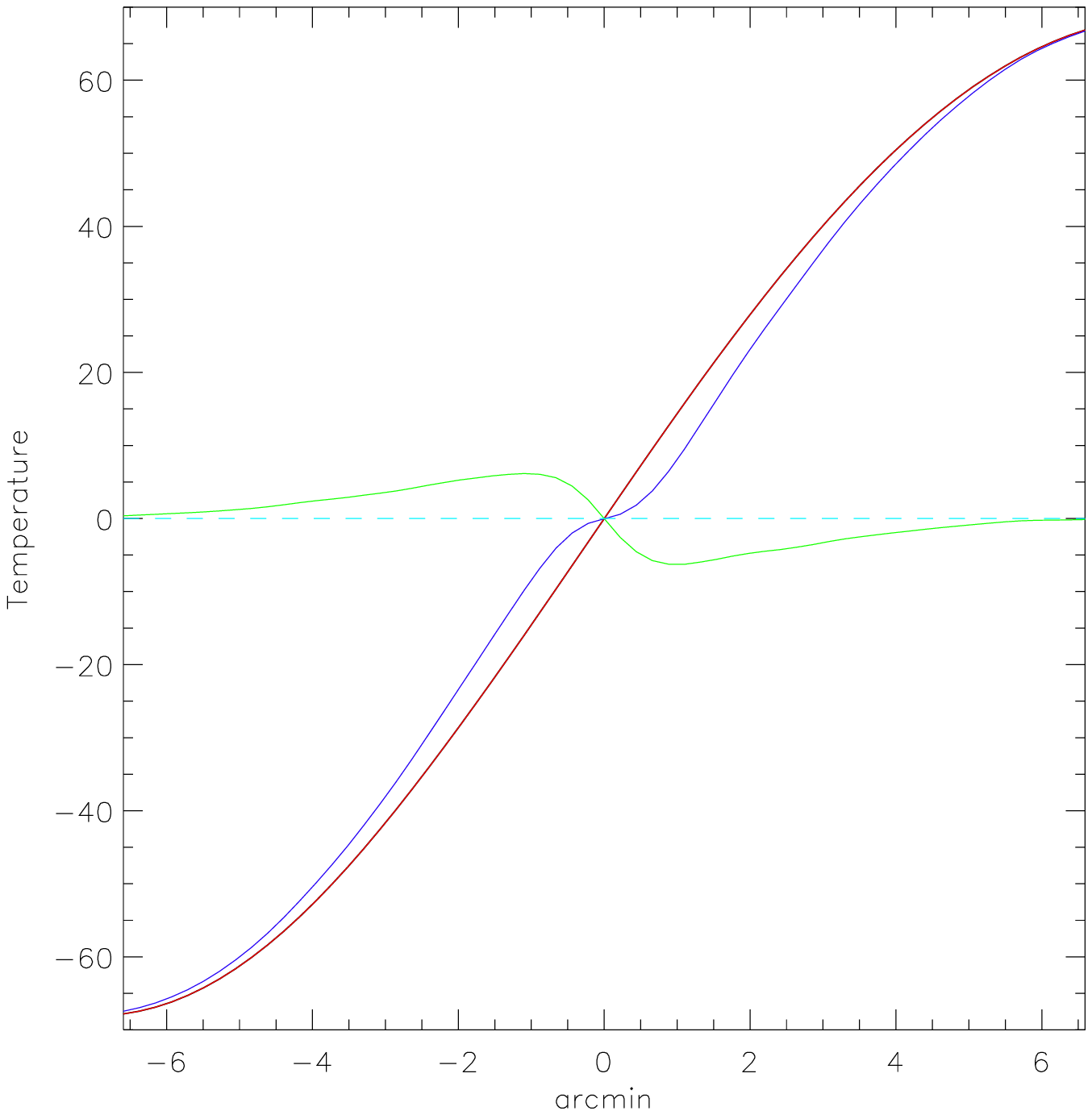}
\includegraphics*[width=7cm]{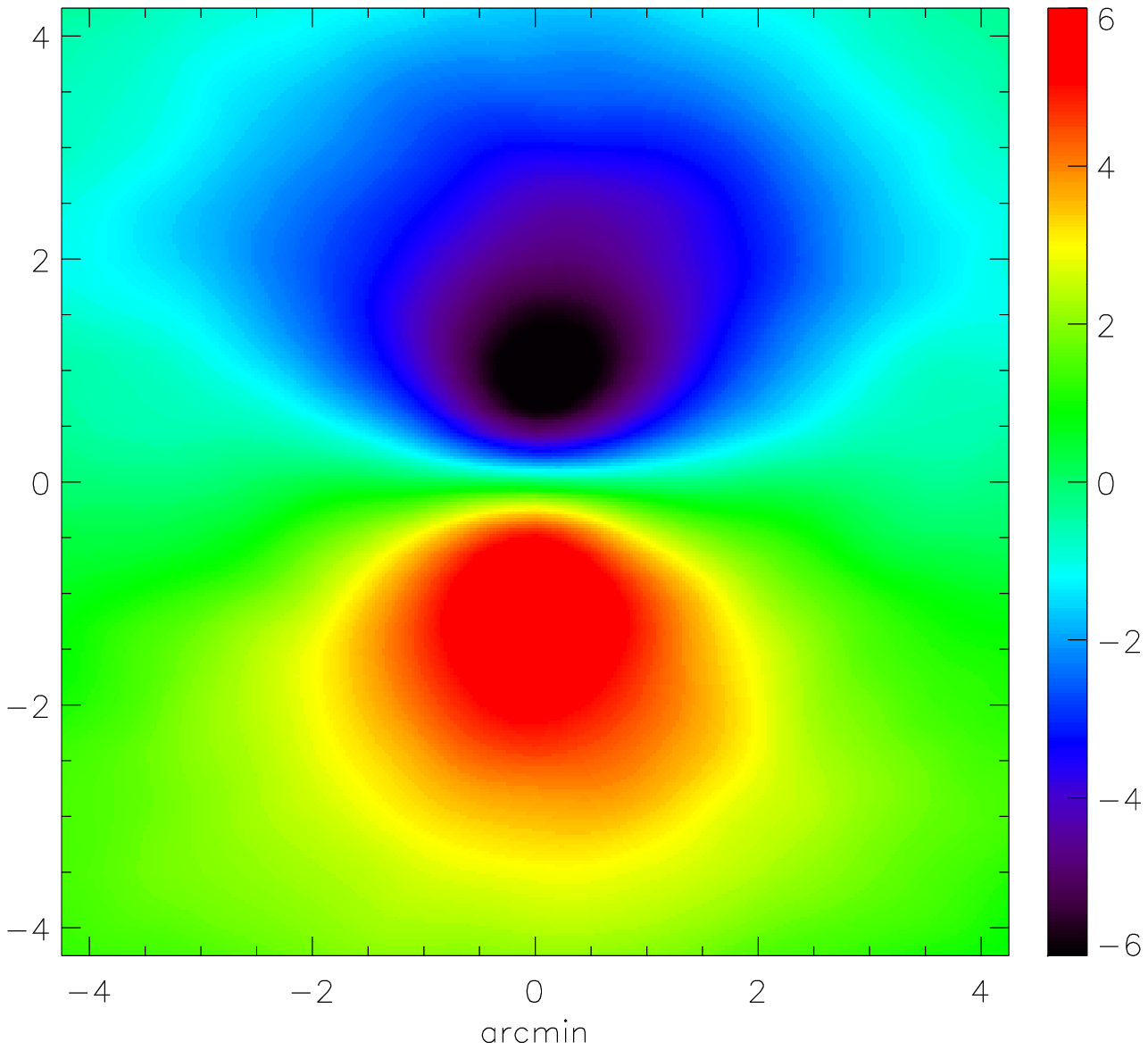}
\end{center}
\caption{(left) A 1-d cut of the lensing signal $\Delta T(\theta)$ for a
circularly symmetric ideal cluster lensing the CMB.  The red line shows
the unlensed CMB, the blue line shows the lensed CMB with the characteristic
`kink' near the origin and the green line shows the difference.
(right) The signal in 2D for a circularly symmetric ideal cluster lensing a
constant gradient.  Note the dipolar nature of the lensing signal.}
\label{fig:lensplot}
\end{figure}

We will assume that equation~(\ref {eq:cmb2}) holds both here and in
\S\ref{sec:toy} for the purpose of illustrating some basic ideas.
However, we note that this approximation is not actually necessary,
and we shall dispense with it altogether in \S\ref{sec:beyond}.  It's
useful to consider the deflection angle due to a spherically symmetric
cluster at comoving distance $\chi$

\begin{equation}
\label{eq:cluster}
\delta \mbox{\boldmath $\theta$} 
  = 4 G {(\chi_s - \chi) \over \chi_s \chi} {M(\theta) \over \theta}
  \mbox{\boldmath $\hat \theta$}
\end{equation}
 
where {\boldmath $\theta $}~$= (\theta_x, \theta_y)$ is now defined as
the angular displacement from the center of the cluster, $\theta$ is
the absolute value of {\boldmath $\theta$}, and $M(\theta)$ is the
mass of the cluster within a radius $\theta$.  To first order, the
cluster's lensing effect is to remap the CMB radially away from its
center, creating a step like wiggle in the CMB gradient centered on
the cluster \cite{SZ00}.  We give an example of this behavior for a
typical large cluster from our simulations in
Figure (\ref{fig:lensplot}).  As expected from
equation~(\ref{eq:cmb2}), the magnitude of the effect is proportional
to both the local gradient of the CMB and the deflection angle along
that gradient.

The lensing angle $\delta${\boldmath $\theta $} cannot be solved for
using equation~(\ref{eq:cmb2}) without more information; you can't in
general measure a scalar field and expect to solve for a scalar field
and a vector field!  If progress is to be made, some assumptions must
be made about $\delta${\boldmath $\theta $}, $\nabla
\tilde{T}$({\boldmath $\theta $}), or both.  We begin by noting that,
in the absence of secondary anisotropies, the CMB is expected to have
little power on small angular scales, so that $ \nabla
\tilde{T}$({\boldmath $\theta $}) may be slowly varying in relevant
regions near the cluster's center.  We examine this statement more
carefully below, but for now we follow \citeasnoun{SZ00} and make use
of this idea to model the primordial CMB
gradient in small regions near the cluster as a constant whose
direction of steepest ascent can without loss of generality be taken
as the y-axis, so that

\begin{equation} \label{eq:grad}
\tilde{T}(\mbox{\boldmath $\theta$}) \approx \tilde{T}_{y 0} \ \theta_y
\end{equation}

where $\tilde{T}_{y 0}$ is the slope of the primordial CMB along the y-axis and we have ignored an overall constant.  

We define the lensing signal due to a cluster $\Delta T$({\boldmath $\theta $}) as the difference between the lensed and unlensed CMB temperature

\begin{equation} \label{eq:signal1}\Delta T (\mbox{\boldmath $\theta$}) \equiv 
T(\mbox{\boldmath $\theta$}) - \tilde{T}(\mbox{\boldmath $\theta$}) \end{equation} 
Combining this with equations~(\ref{eq:cmb2}) and (\ref{eq:grad}) and defining the deflection due to lensing by the cluster along the y-axis as $\delta \theta_y $ gives

\begin{equation} \label{eq:signal2}
\Delta T (\mbox{\boldmath $\theta$}) \approx
T(\mbox{\boldmath $\theta$}) - \tilde{T}_{y 0} \theta_y \approx 
- \delta \theta_y(\mbox{\boldmath $\theta$}) \ \tilde{T}_{y 0}
\end{equation}

In Figure (\ref{fig:lensplot}) we show $\Delta T$({\boldmath $\theta $})
for a circularly symmetric isolated cluster lensing a constant
gradient.  The signal crudely resembles a dipole in appearance as you
would expect from equation (\ref{eq:cmb2}), though it falls off as
$\sim M(\theta) / \theta$ away from the cluster as predicted by
equation (\ref{eq:cluster}).

The lensing angle $\delta \theta_y$({\boldmath $\theta $}) is not
generally considered measurable because the original position of the
background image isn't known.  However, far away from the cluster, the
effect of lensing must be small, and the lensed and unlensed CMB must
be roughly equal.  If the approximation of equation~(\ref{eq:grad}) is
valid at this distance, then it will be possible to measure
$T$({\boldmath $\theta $}) away from the cluster, determine
$\tilde{T}_{y 0}$, and solve for $\delta \theta_y$({\boldmath $\theta
$}).  As we show in Figure (\ref{fig:disaster}), the convergence profile
can be well reconstructed from this information.  We note that the
reconstruction is degenerate for density fluctuations that change
$\delta \theta_x$ but don't alter $\delta \theta_y$.  This degeneracy
is similar to that of the mass sheet degeneracy, but it applies to any
line of constant density in $\Sigma$({\boldmath $\theta$}) that
happens to lie in the direction of the y-axis.  We find the error due
to this degeneracy to be quite small, and it is of course identically
zero for a circularly symmetric cluster profile, where $\delta
\theta^2 = \delta \theta_x^2 + \delta \theta_y^2$

It is clear from Figure (\ref{fig:disaster}) that reconstructing the convergence profile of a cluster from CMB temperature maps is certainly possible under the following highly artificial conditions:  

\begin{itemize}\item No foregrounds
\item No instrument effects
\item No CMB secondary anisotropies other than lensing
\item The clusters are isolated
\item The CMB is a pure gradient of constant slope\end{itemize}

In \S\ref{sec:toy}, we begin to include these effects in the context of a toy model.

\section {Results From a Toy Model} \label{sec:toy}

In this section, we make use of a toy model of cluster lensing of the
CMB in order to investigate to what extent the issues raised in the
bulleted points listed at the end of \S\ref{sec:theory} will impact on
our ability to reconstruct the convergence profiles of clusters using
high resolution CMB temperature maps.  We then present results from
this toy model for two general cases: individual clusters and
``stacked'' average clusters.

\subsection {Signal and Noise in the Toy Model} \label{sec:toysignal}

In the toy model, we assume that the primordial CMB in a small region
near a cluster is a known quantity, which we then model as a gradient.
This allows us to bypass the step of estimating the unlensed CMB from
the actual maps (we address this issue in \S\ref{sec:beyond}), which
will be both lensed and noisy, and instead to directly measure the
signal as defined in equation~(\ref{eq:signal1}) plus a noise term
$N$({\boldmath $\theta$}), which includes all other effects, so that

\begin{equation} \label{eq:signal3}\Delta T_t (\mbox{\boldmath $\theta$}) \equiv 
\Delta T (\mbox{\boldmath $\theta$}) + N(\mbox{\boldmath $\theta$})\end{equation}

where we define the measured signal in the toy model as $\Delta T_t$.
According to equation~(\ref{eq:cluster}), the deflection angle $\delta
\theta \sim M(\theta) / \theta$, and from equation~(\ref{eq:signal2})
the signal in any circular annulus is proportional to $\delta \theta$
times the slope of the CMB gradient $\tilde{T}_{y 0}$ times the area
of the annulus.  Then the total signal to noise inside a circle of
radius $\theta$ scales as

\begin{equation} \label{eq:scale}
{S \over N} \sim {\tilde{T}_{y 0} \over \theta^{1 \over 2 }} \int d \theta M(\theta)
\end{equation}

for Gaussian noise.  Since the model relies on the approximation that
the CMB gradient is constant (equation~\ref{eq:grad}), which can only
be valid on small scales, we have chosen to impose a cut-off radius of
$4'$ from the center of the cluster.  That is, we assume a measurement
of $\Delta T_t$ for $\theta < 4'$, and use no information at larger
radii.  We note that we do address the use of lensing information for
reconstruction on larger scales elsewhere \cite{AVW03}.

The toy signal $\Delta T_t$ is derived using an input convergence map
made by ray-tracing through our N-body simulation.  This convergence
map is then used to make the deflection field $\delta \theta_y$ using
Fourier methods, and $\Delta T$ is made by remapping $\tilde
T$({\boldmath $\theta$}$'$) into $T$({\boldmath $\theta$}) according
to equation~(\ref{eq:cmb1}), where $\tilde T$({\boldmath $\theta$}$'$)
is a gradient of constant slope.  We provide a description of the
N-body simulation and the creation of the convergence map in an
appendix, and for now note only that the simulated clusters are in
general neither isolated nor ideal, as can be seen in the convergence
profiles of Figure (\ref{fig:clusters}).

After creating a map of the toy signal, we next introduce two noise
components to the model.  The first, the kinetic SZ, is highly
correlated with the position of the cluster, and because it has the
same spectral dependence as the CMB itself, it cannot be removed using
multi-frequency measurements.  We then add a Gaussian ``white noise''
component, which is uncorrelated with the location of the cluster, and
can be thought of as instrument noise, and unless stated otherwise, we
smooth with a $0.'75$ beam (FWHM), such as is expected for APEX-SZ.  
We do not include noise sources which we
expect to be small and uncorrelated with the cluster, such as other
CMB secondaries, nor do we include point sources, dust, or the thermal
SZ, which can in principle be removed or at least reduced by making
use of their spectral dependence.  In any real experiment, treatment
of these effects will not be perfect, so excluding them entirely is
somewhat optimistic.

\subsection {Reconstruction of Cluster Profiles} \label{sec:toyrecon}

In this section we address the issue of the reconstruction of cluster
convergence profiles within the toy model both for individual clusters
and ``stacked'' average clusters.  We shall see that both instrument
noise, and more perniciously the kinetic SZ, significantly degrade the
reconstruction for individual clusters.

\begin{figure} 
\leftmargin=2pc
\begin{center}
\includegraphics*[height=4.6cm]{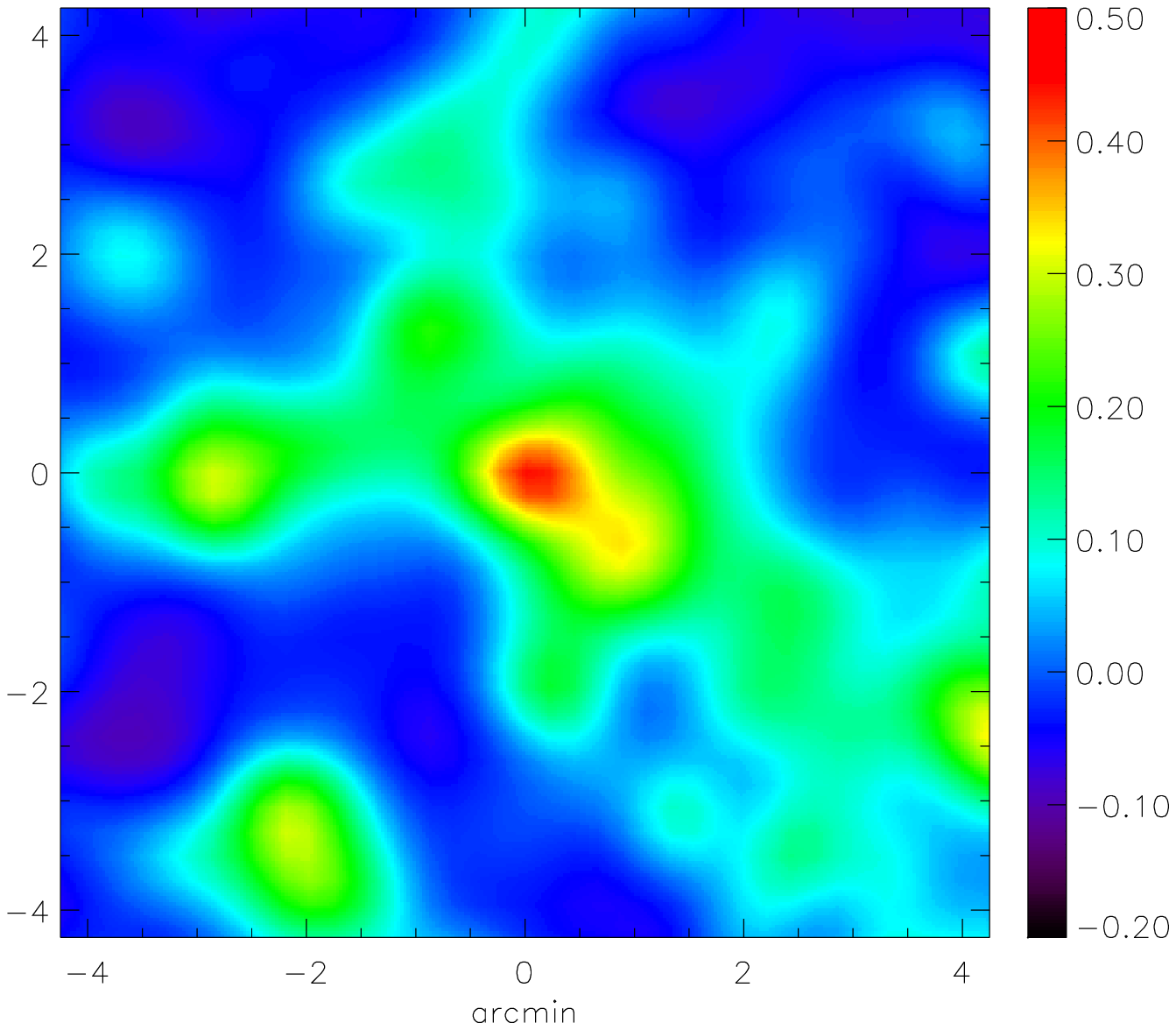}
\includegraphics*[height=4.6cm]{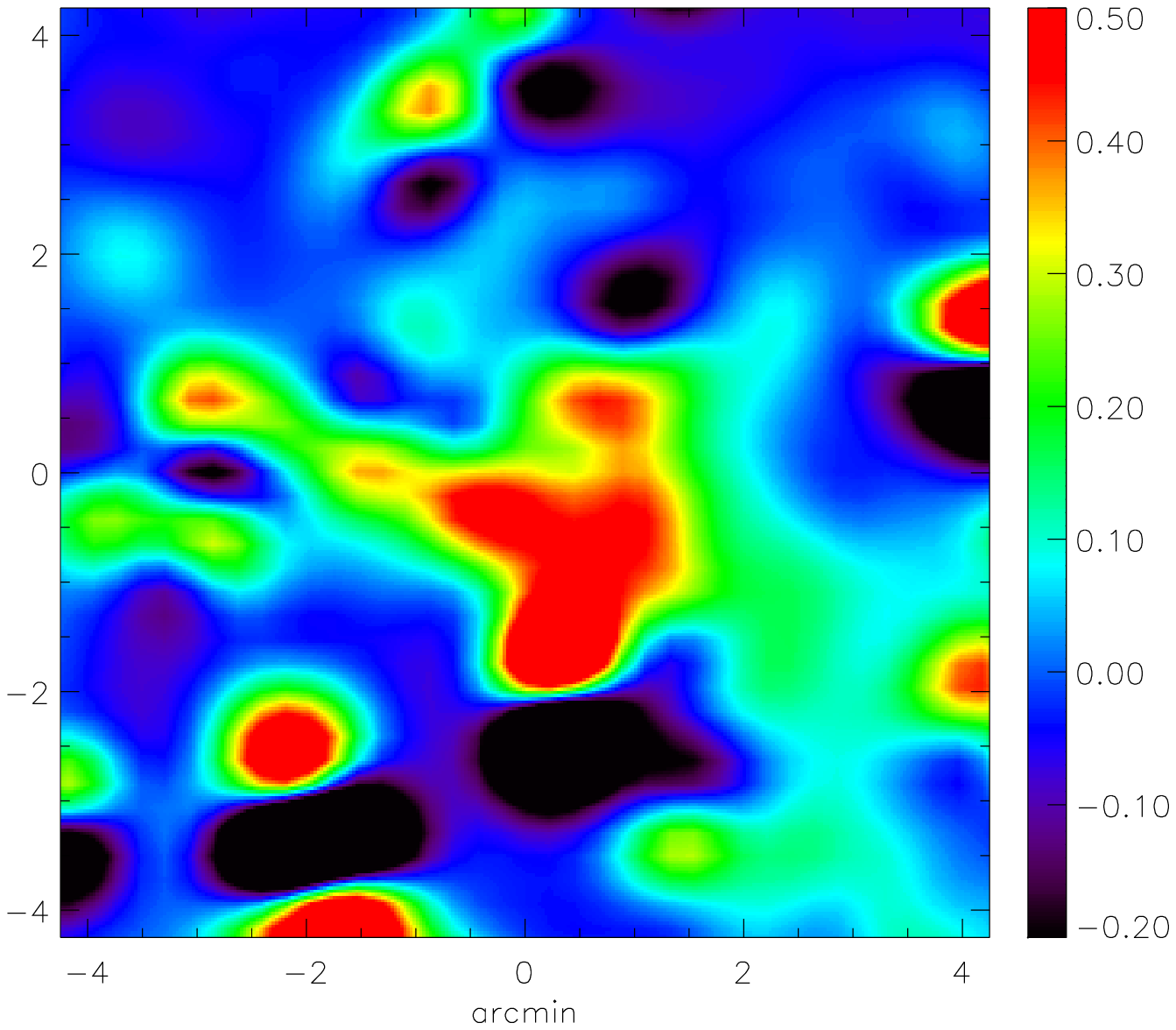}
\includegraphics*[height=4.6cm]{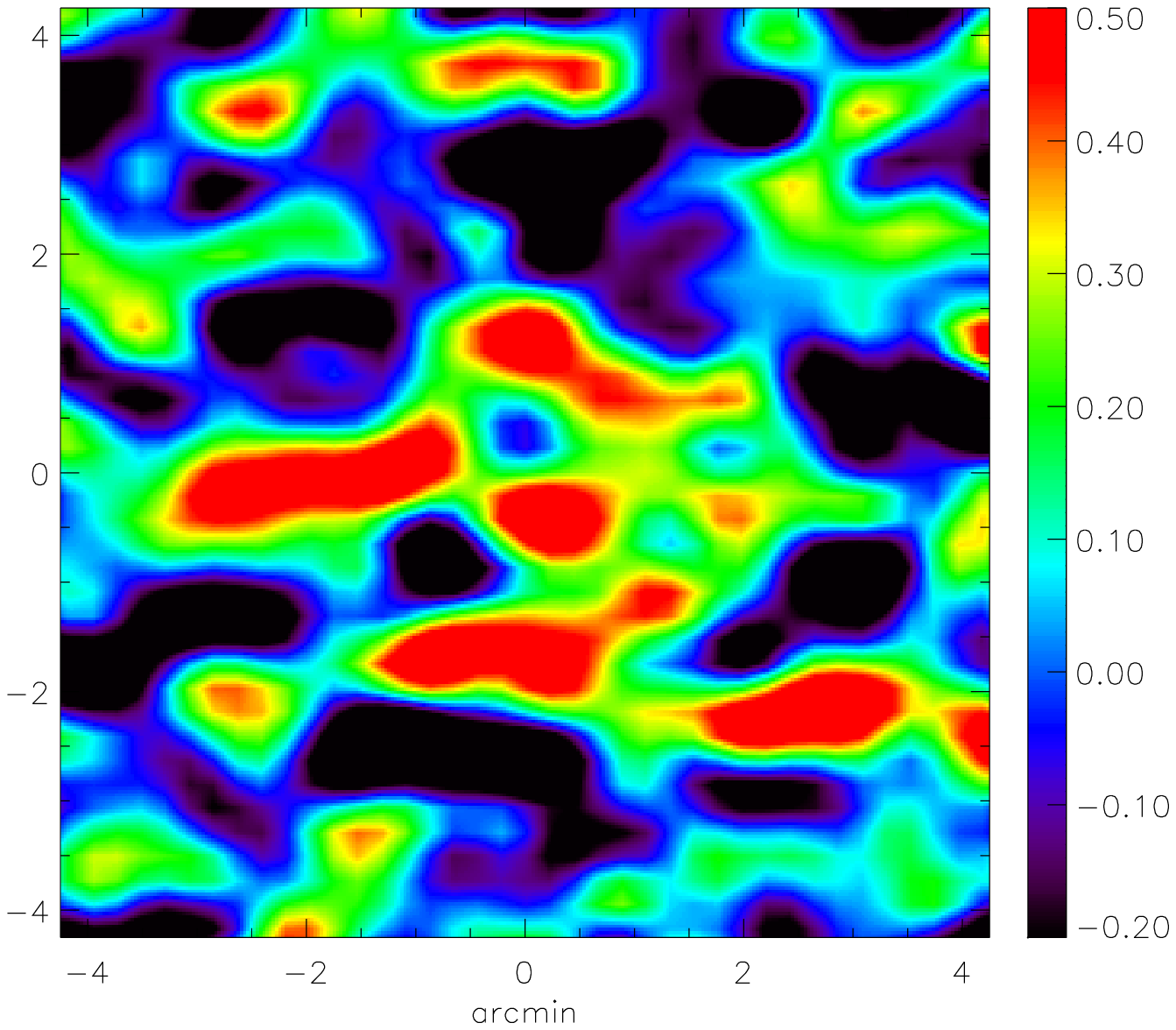}
\end{center}
\caption{The reconstructed convergence maps for a typical cluster
of $M_{200}=2.35\times 10^{14}\,h^{-1}M_\odot$ and $z=1$.  We show
the input map, the reconstruction with kinetic SZ, then with
$3\,\mu$K-arcmin of instrument noise and no kinetic SZ.  }
\label{fig:cluster0}
\end{figure}

Let us begin by looking at the reconstruction including kinetic SZ for a
typical large mass, high-$z$ cluster in Figure (\ref{fig:cluster0}).
Note that when the kinetic SZ is added the reconstruction is significantly
altered, because some of the kinetic SZ signal is misinterpreted as lensing.
This effect is particularly troublesome because it is spatially
correlated with the lensing signal, spectrally indistinguishable from
it, non-Gaussian, and only loosely correlated with the mass or thermal
SZ signal from the cluster.

In the last panel of Figure (\ref{fig:cluster0}), we show a
reconstruction with no kinetic SZ but now instead including Gaussian
random noise at a level of $3\,\mu$K-arcmin, roughly one third of
that expected for the highest resolution maps from APEX-SZ and
consistent with more ambitious future projects such as ACT and SPT.
Even at this noise level, most of the features of the convergence map
are lost, and it is evident that high quality reconstruction of
cluster convergence maps for typical clusters is not achievable within
the context of the experimental parameters we are considering.  A
strategy designed to integrate deeply on cluster locations would have
to be adopted.

\begin{figure} 
\leftmargin=2pc
\begin{center}
\includegraphics*[height=4.6cm]{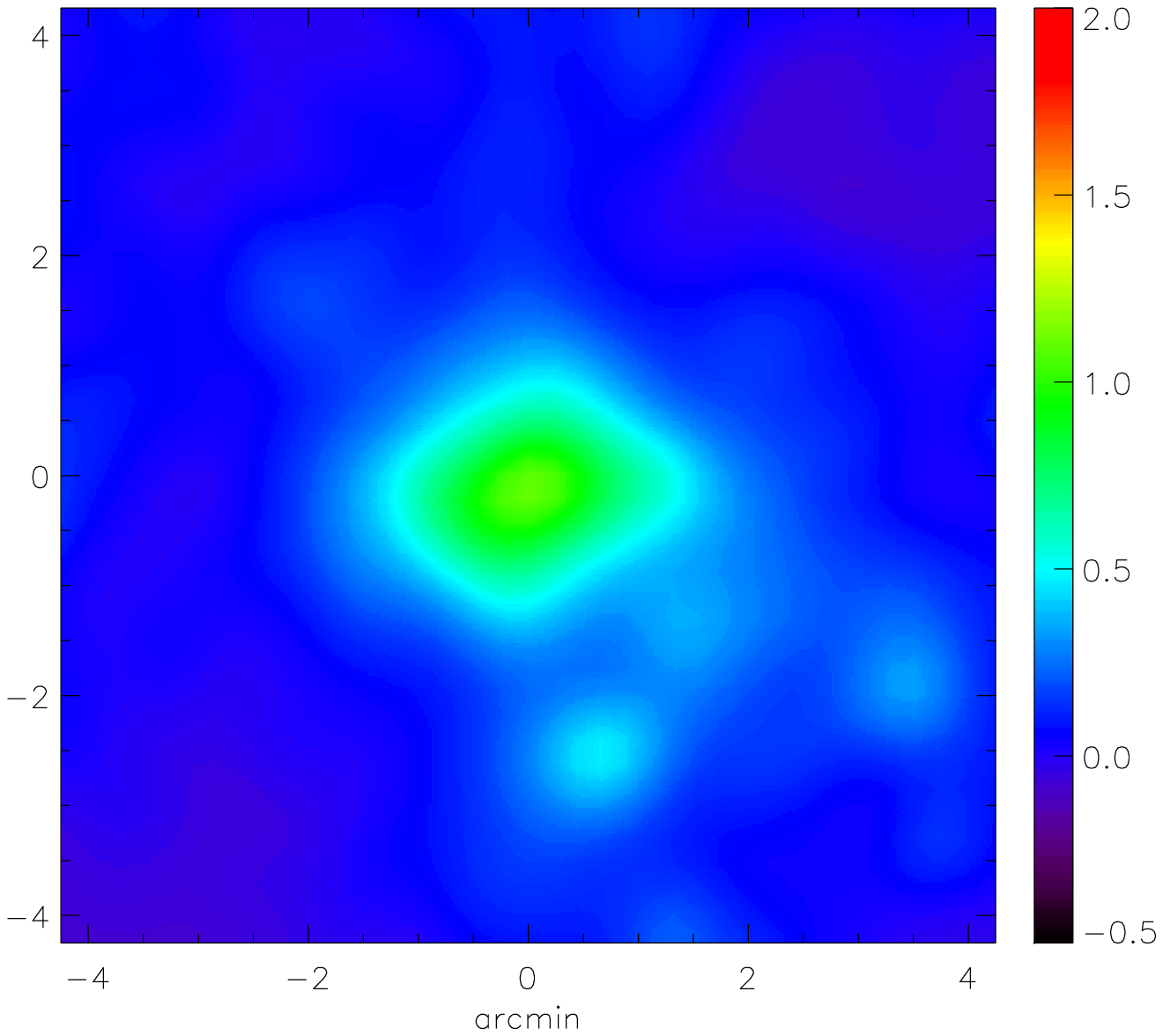}
\includegraphics*[height=4.6cm]{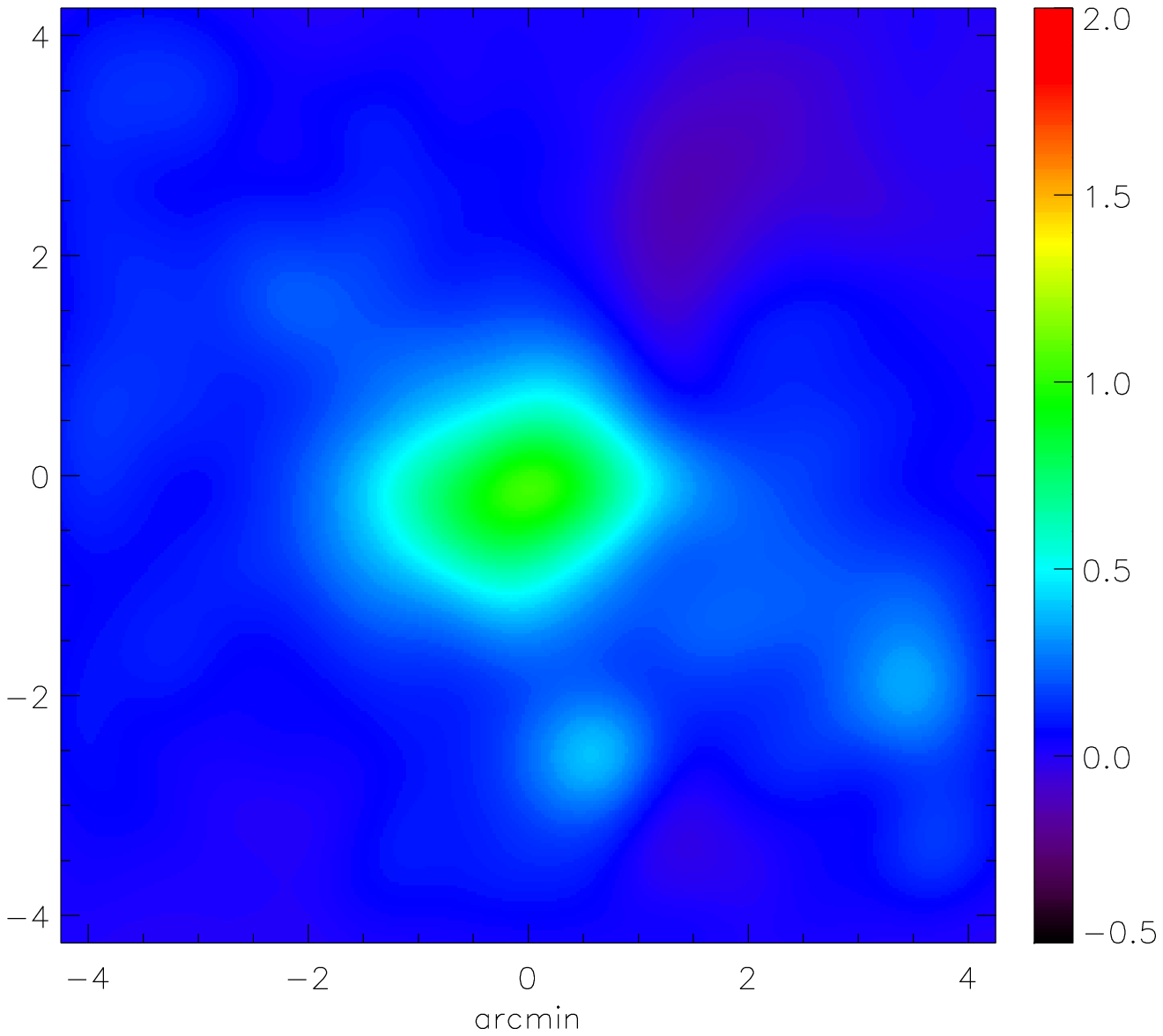}
\includegraphics*[height=4.6cm]{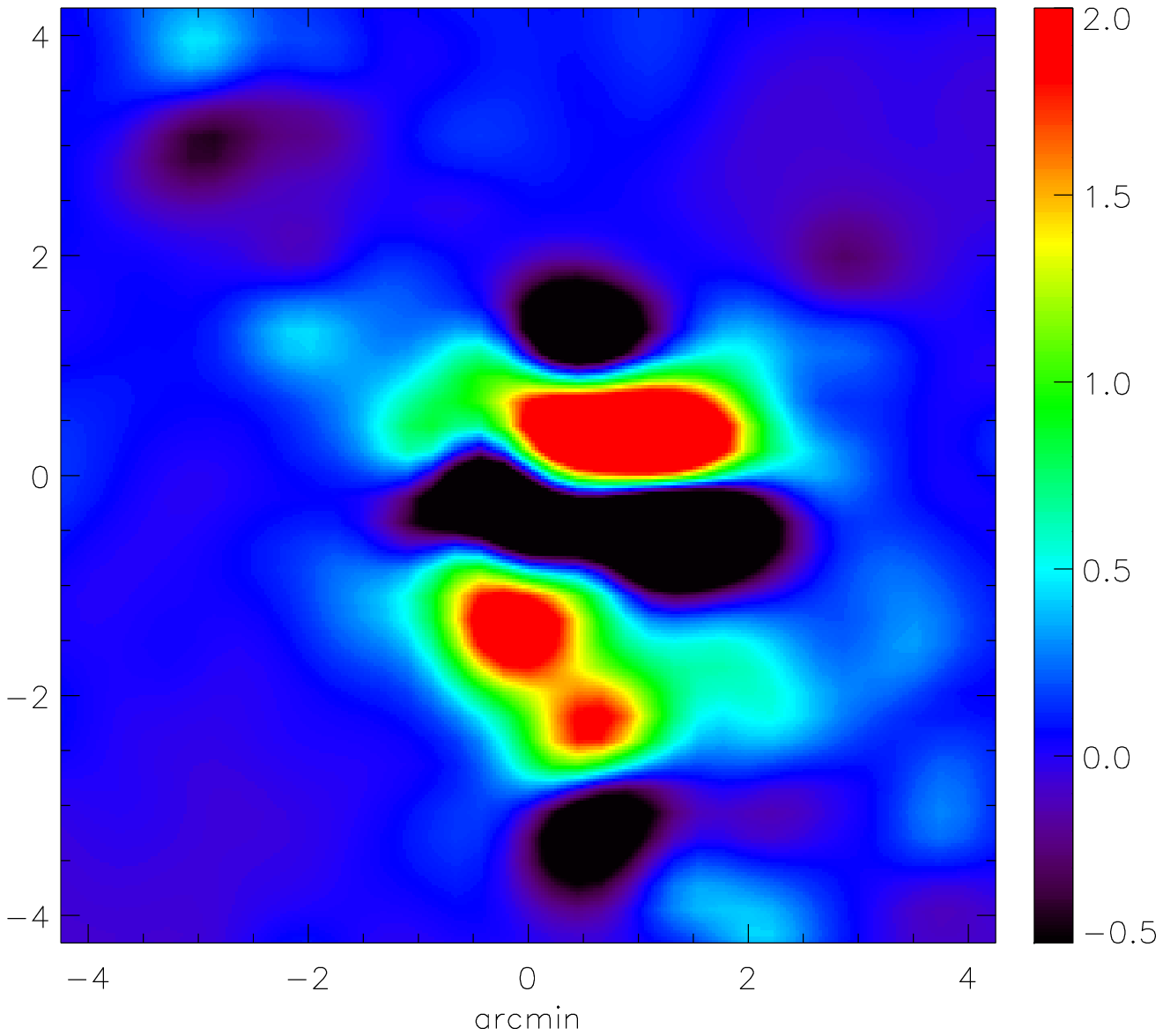}
\end{center}
\caption{The convergence reconstruction for a large
($M_{200}=7\times 10^{14}\,h^{-1}M_\odot$), high redshift ($z=0.95$) cluster.
The left panel is the true convergence smoothed by $0.'75$, the center panel
is the the reconstructed convergence using the toy model and no noise,
and the right panel is the reconstructed convergence including the kinetic SZ.
In this unusual case, the kinetic SZ dominates the reconstruction completely,
causing the center of the cluster to appear as large and negative.}
\label{fig:disaster}
\end{figure}

While the situation depicted in Figure (\ref{fig:cluster0}) is
typical, in some cases the kinetic SZ can be completely dominant, as
we show in Figure (\ref{fig:disaster}).  In this particular case the
cluster happens to be rotating, so that the kinetic SZ signal has a
dipole-like structure similar to that produced by lensing.  Depending
on the relative orientation of the kinetic SZ lobes and the direction
of the CMB gradient, these signals can enhance or overwhelm the
lensing signal.  It is even possible to reconstruct a large negative
convergence right at the cluster's location!

\begin{figure} 
\leftmargin=2pc
\begin{center}
\includegraphics*[height=4.4cm]{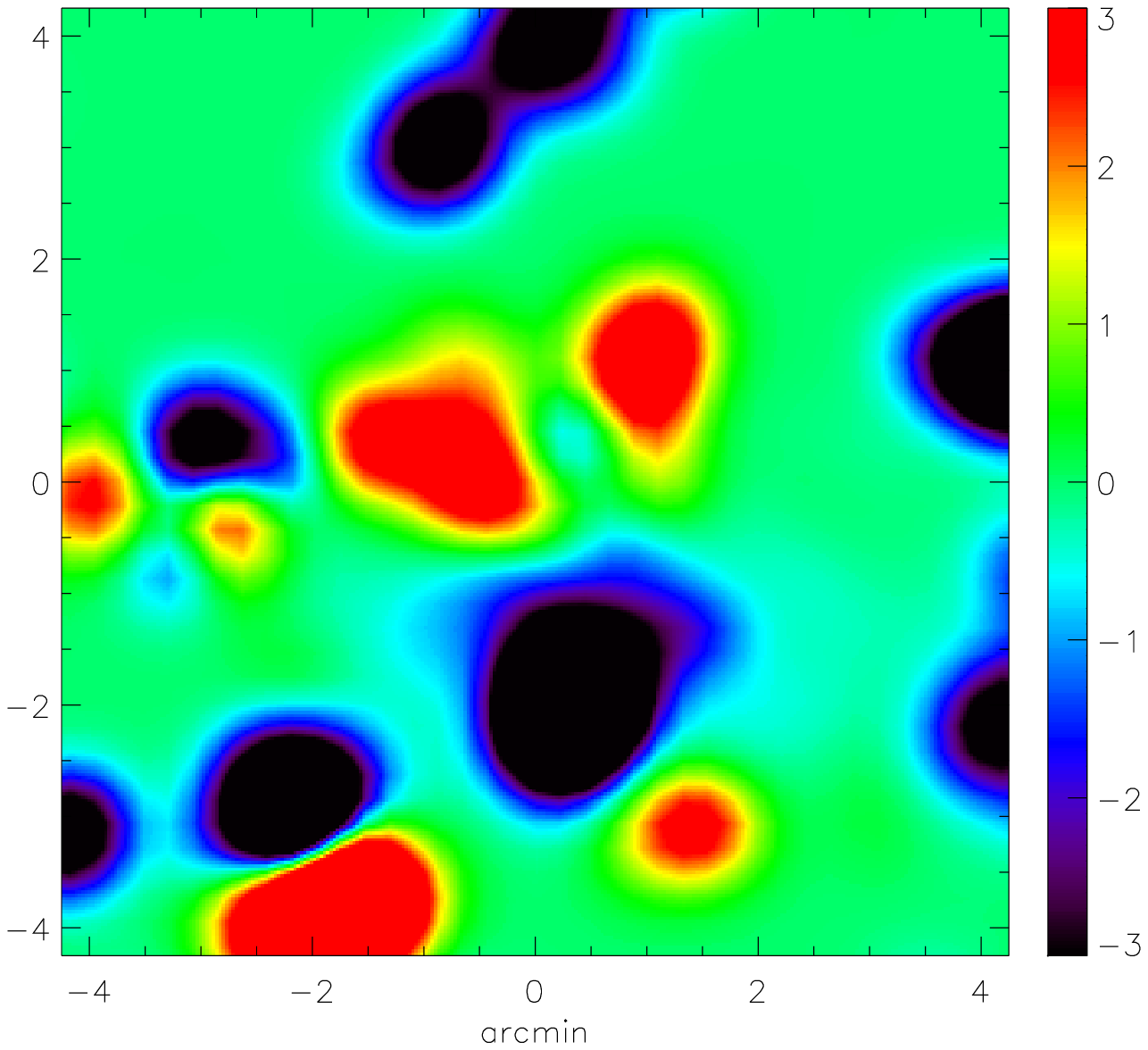}
\includegraphics*[height=4.4cm]{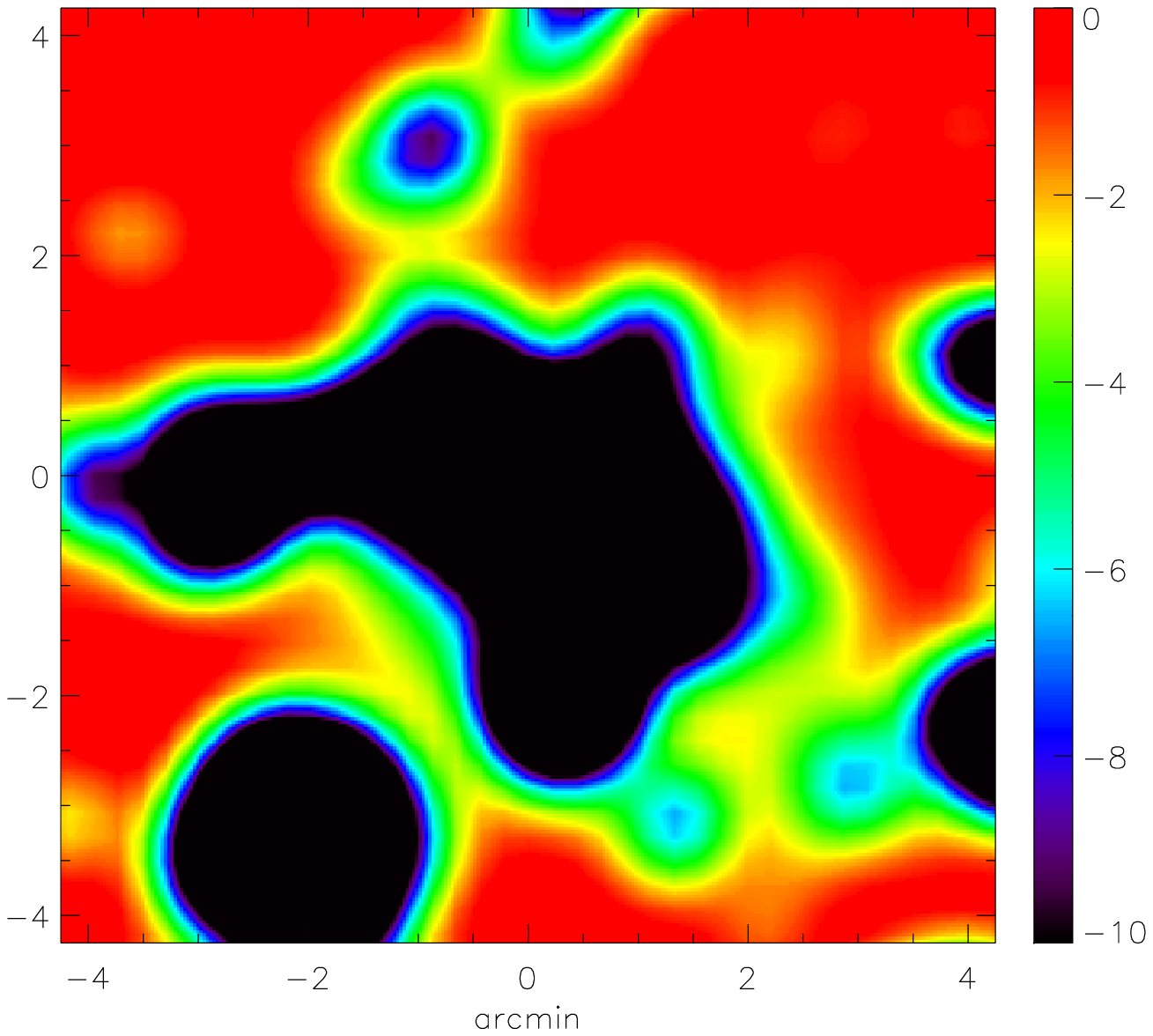}
\includegraphics*[height=4.4cm]{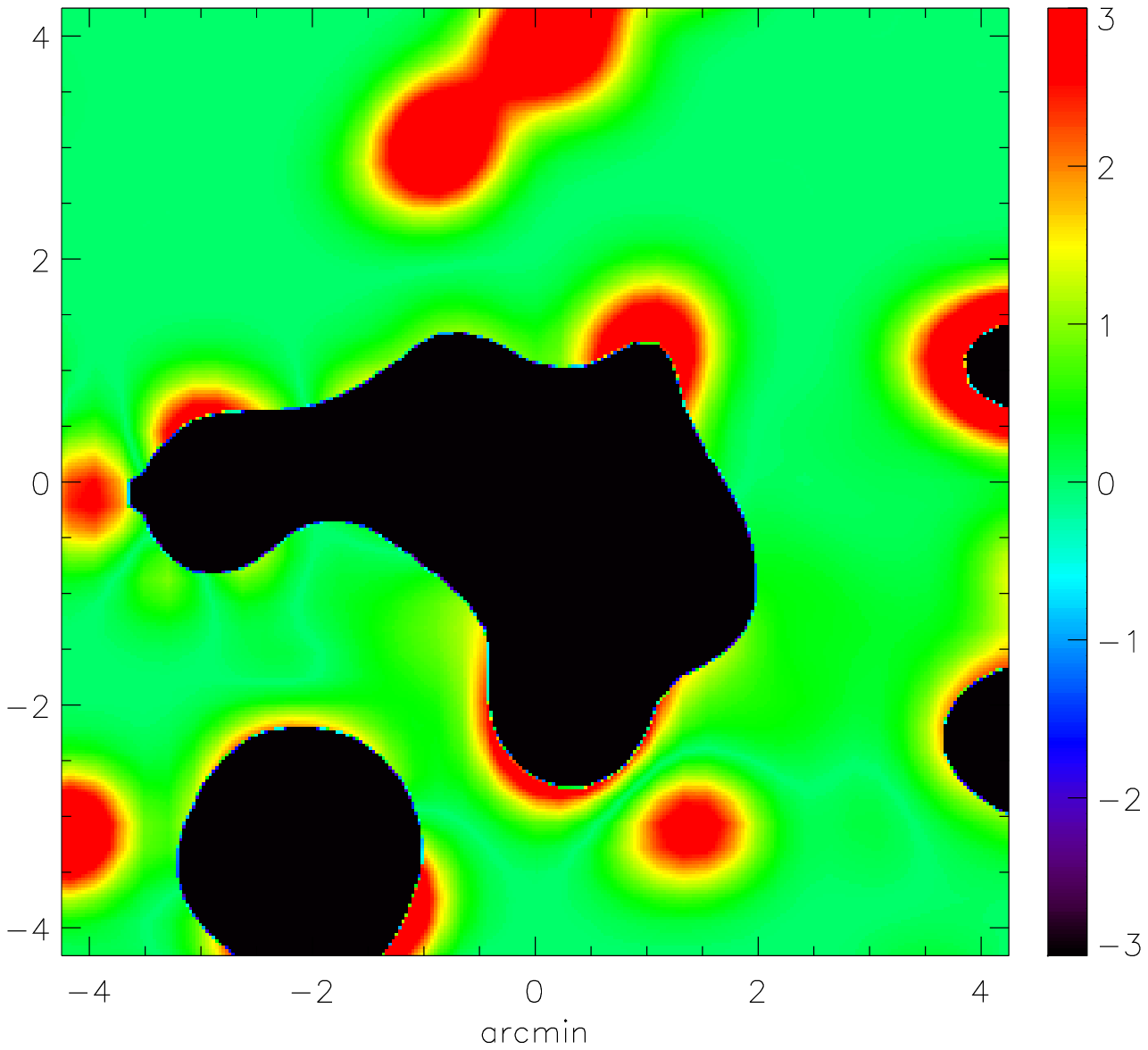}
\end{center}
\caption{The kinetic SZ (left) for a typical (non-isolated) cluster and the
thermal SZ (center) that you could try to use to mask it.  The right panel 
is an example of a typical masked map, where the masked region is shown 
in black and the unmasked region is now (for clarity) the absolute value of 
the kinetic SZ.  The thermal SZ is not a perfect tracer of the kinetic SZ, resulting 
in map with many perfectly good pixels removed, and residual kinetic SZ 
still included.  The color bar in each case shows $\Delta T$ in $\mu$K. }
\label{fig:tszksz}
\end{figure}

While this ``disaster'' cluster is clearly beyond hope, the
reconstruction of a typical cluster might be improved if we mask out
pixels which are likely to contain a large kinetic SZ component, which
can be done by using the thermal SZ to roughly estimate the likelihood
of a large magnitude kinetic SZ.  This is essentially the idea
proposed in \citeasnoun{SZ00}.  Figure (\ref{fig:tszksz}) illustrates
why this is not in general helpful; most clusters are not isolated,
and many pixels must be excised.  To make matters worse, the thermal
SZ is not a perfect tracer of the kinetic SZ, so many perfectly good
pixels are thrown away while others with large kinetic SZ signals are
included.

Obviously, even if most clusters are not suitable candidates, we may
be able to select some that are and focus on those.  We have so far
been thwarted in our reconstruction efforts by the kinetic SZ and
instrument noise.  If we ignore the latter for the time being, we
realize right away that, although rare, clusters do sometimes form in
relative isolation, as depicted in Figure (\ref{fig:clusters}).
Perhaps these will be suitable?

Indeed, the level of contamination from the kinetic SZ is dramatically
lower, and mostly associated with the cluster itself.
We may now mask out the center of the cluster and have a reasonable
expectation that the kinetic SZ contamination will be under control,
and indeed that does prove to be the case here.
However, we have specifically gone out of our way to select an isolated
cluster, using the thermal SZ, and this has introduced a strong selection
bias.
The absence of other structures is marvelous for controlling the impact
of the kinetic SZ, but the cluster is now located in a large void which
dramatically reduces the projected mass below what one would expect from
an average location.  In essence we have introduced a mass sheet degeneracy.
This is depicted in Figure (\ref{fig:deflection}).
Here, we present the actual deflection angle along the y-axis.
The dot-dash curve is the lensing angle that would result if the
cluster were the only object in the universe, the solid line is the
result including the cluster plus projection effects, and the dashed
line is the actual deflection angle, where structures far away from
the cluster of interest contribute to the deflection angle even though
they don't alter the convergence (this last is a degeneracy that arises
because we have access only to the y-axis deflections).  Thus, even though
we ``cheated'' and simply used the y-axis deflection angle directly, rather 
than measuring it, we are still unable to reconstruct the clusters mass from 
this information due to a strong mass sheet degeneracy that in fact is 
consistent with a cluster mass estimate of zero at the virial radius, 
with absolutely no noise introduced, simply because of the confusion 
introduced from lensing by other structures.

\begin{figure} 
\leftmargin=2pc
\begin{center}
\includegraphics*[width=10cm]{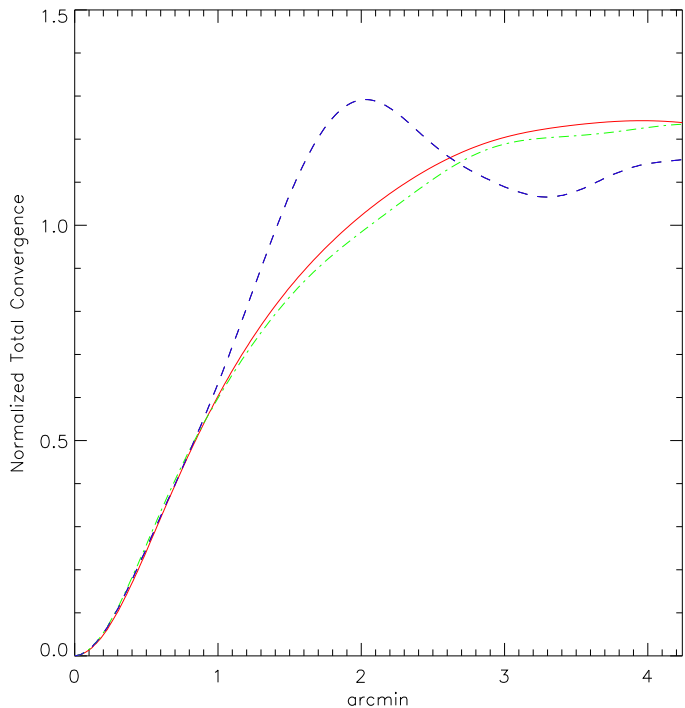}
\end{center}
\caption{The cumulative convergence for our stack of 61 clusters with
$2<M/10^{14}\,h^{-1}M_\odot<3$ and $1<z<1.5$ as a function of radius.
Shown here are the input (red line),
the reconstruction including kinetic SZ (green line),
and reconstruction including kinetic SZ and noise (blue line).}
\label{fig:profile}
\end{figure}

Our investigation of the likely success of reconstruction of individual
cluster mass profiles has shown that it will prove a more difficult problem
than one might have hoped.  It is worth noting that some of the difficulties
we have encountered have come because we have used full field lensing and
SZ simulations, rather than clusters that have been `cut out' of
simulations and then used to lens voids.
Despite this technical difference though, our results so far are
qualitatively in agreement with those of \citeasnoun{HK04}.
They show in their Fig.~4 that they reconstruct the cluster mass to be
essentially zero at the virial radius.  We shall return to this point,
and some of the reasons behind it, in the next section.

\begin{figure} 
\leftmargin=2pc
\begin{center}
\includegraphics*[height=4.9cm]{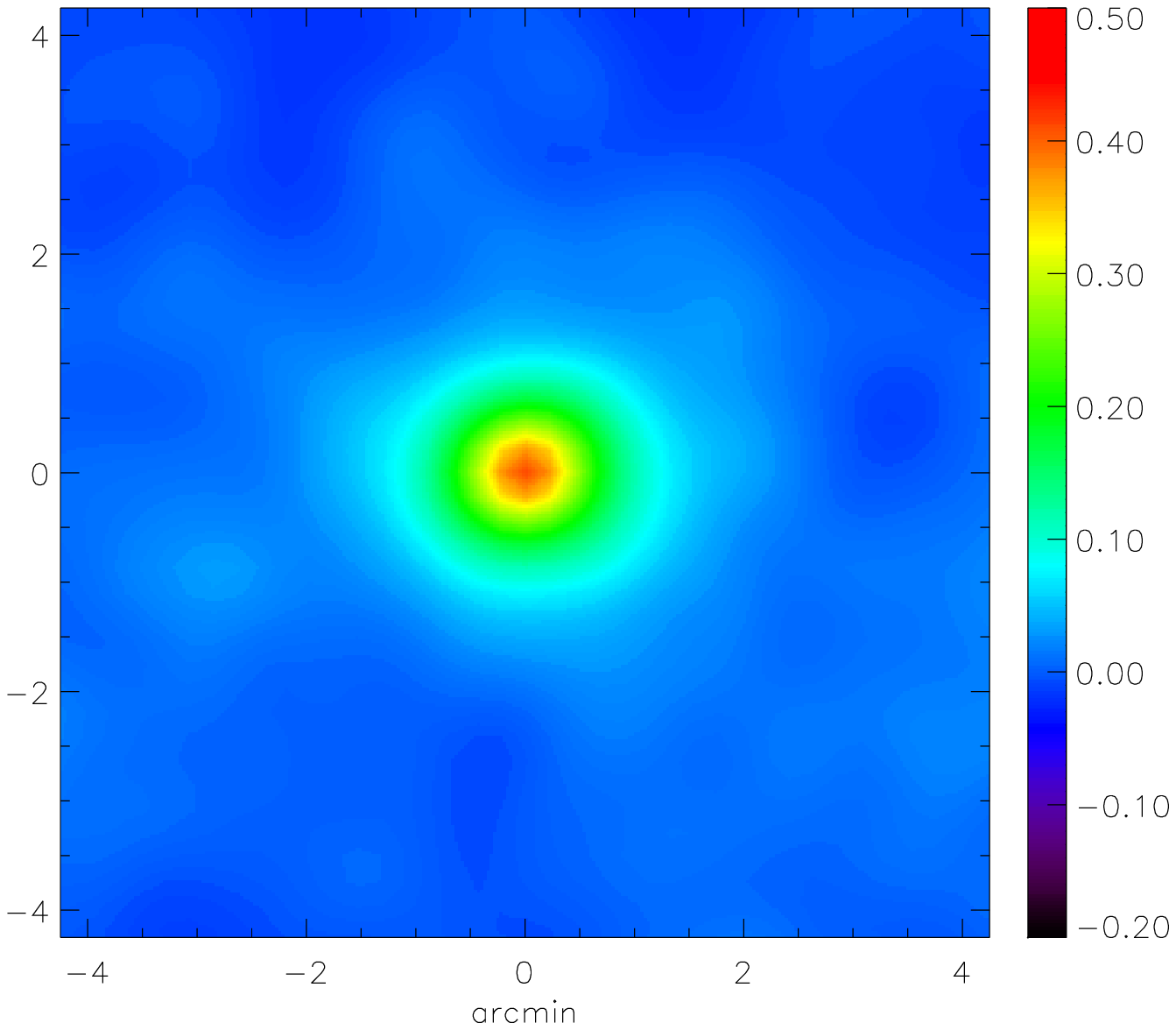}
\includegraphics*[height=4.9cm]{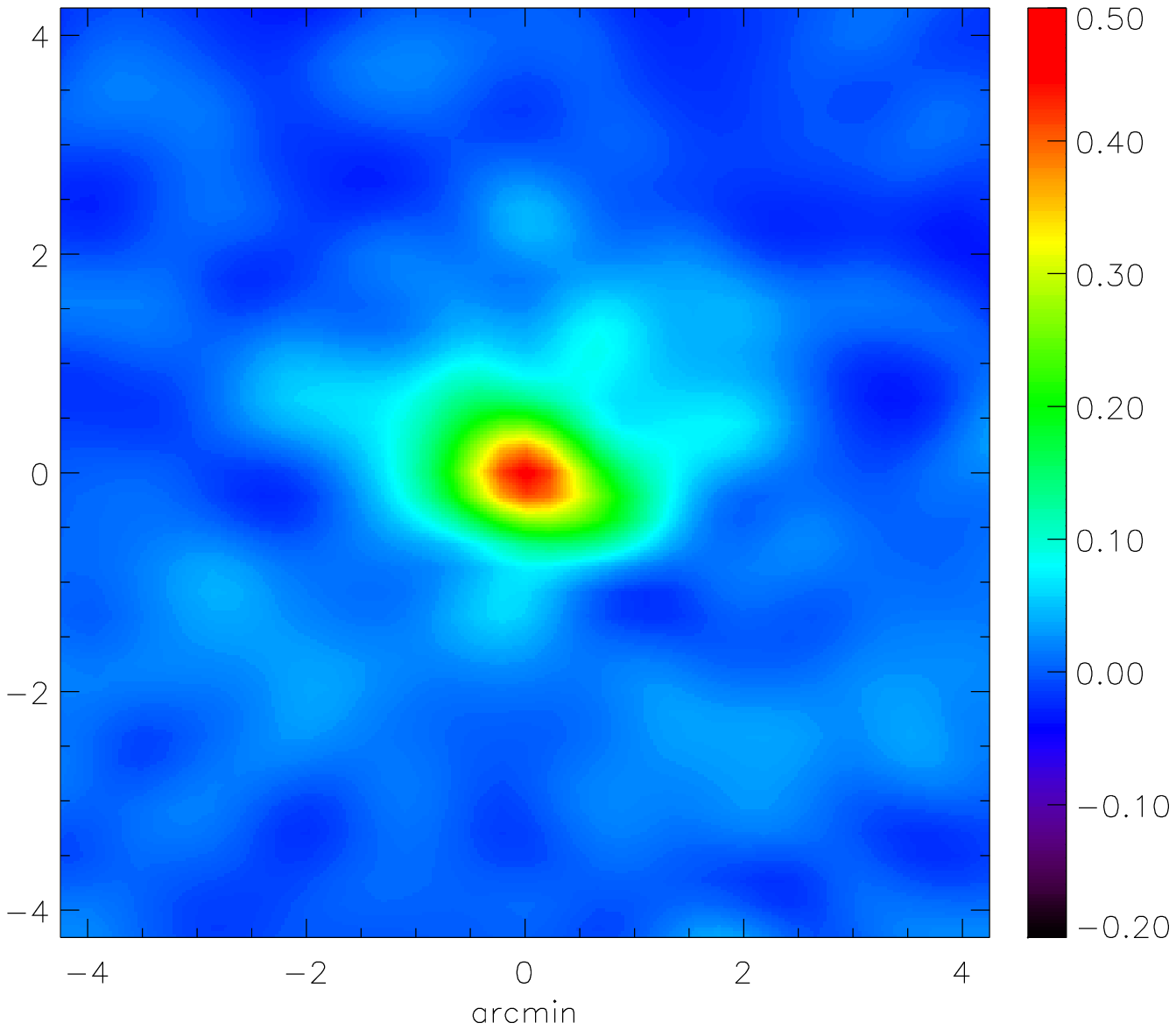}
\includegraphics*[height=4.9cm]{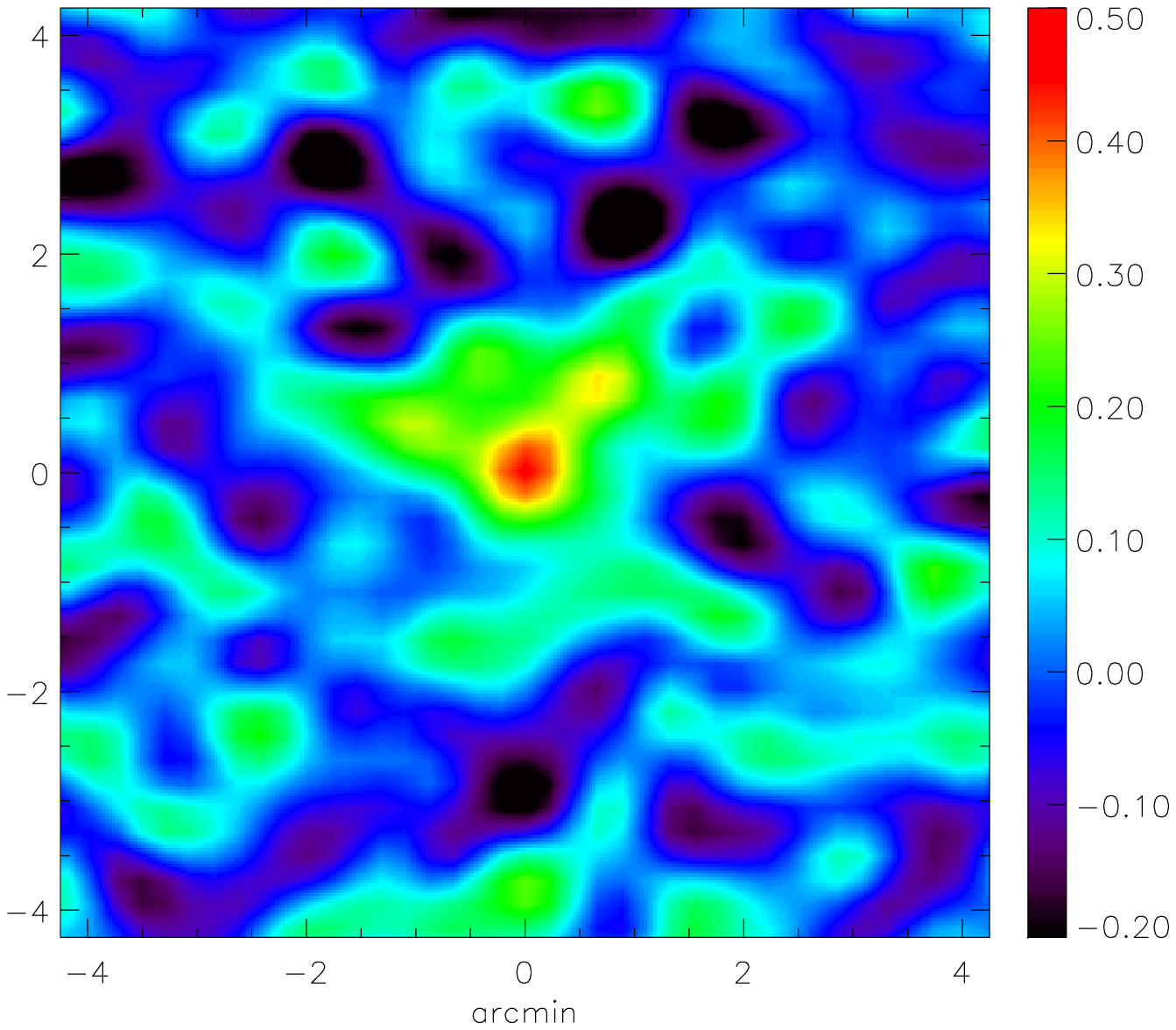}
\end{center}
\caption{The convergence map (left) of our stack of 61 clusters with 
$2<M/10^{14}\,h^{-1}M_\odot<3$ and $1<z<1.5$, it's reconstruction (center) 
including the kinetic SZ, and then again (right) with both Gaussian noise and kinetic
SZ included.  }
\label{fig:stack}
\end{figure}

Another approach, called stacking, is to consider a group of clusters
binned according to relevant observables (e.g.~temperature and
redshift), cut out regions in the signal maps within a specified area
centered on each cluster, stack them one on top of another, and then
compute the average.  The principal advantage of this technique is
that essentially every unbiased source of confusion is reduced by this
kind of averaging as $\sqrt{\rm N}$, where N is number of clusters
used.  This includes the important cases of the kinetic SZ, since the
baryons are a priori equally likely to be moving toward or away from
the observer, and of projection effects that proved so troublesome in
the case of ``isolated'' clusters.
If our cluster signal grows as $M$, our signal-to-noise is enhanced by
stacking more relatively low mass clusters as long as the mass function
is steeper than $M^{-2}$ per mass interval considered.
This is the case for the high mass and redshift clusters we consider,
suggesting we should target more lower mass clusters rather than fewer
high mass clusters.

In Figure (\ref{fig:stack}) we show the reconstruction of a stack made
{}from 61 intermediate mass, high redshift clusters.  First, consider
the case where the only source of noise is the kinetic SZ.  The convergence
profile of the stack is circularly symmetric to a good approximation,
which greatly reduces the mass line artifact that can arise in the
reconstruction of finite fields, and it becomes feasible to
reconstruct more of the central region.  As in the case of single
clusters, adding Gaussian instrument noise of $1.25 \ \mu$K-arcmin to
the stack (consistent with the anticipated $10 \ \mu$K-arcmin for
APEX-SZ averaged over 61 clusters) introduces substantial uncertainty
to the convergence profile, but the total convergence as a function of
radius is reasonably constrained, as can be seen in
Figure (\ref{fig:profile}).  Thus, there is some promise that
one might be able to provide reasonable mass estimates for cluster
stacks, which might then be used to constrain the mass-temperature
relation.

The reconstruction of cluster convergence profiles using weak lensing
of the CMB is a technique that clearly shows promise for cluster
stacks if the approximations of the toy model can be trusted.  Some
potentially important issues are not addressed by this simple model,
the most important of which is the fact that the unlensed CMB is not a
known quantity, so the signal as described in
equation~(\ref{eq:signal2}) is not directly measurable.  We examine
this issue in the next section.

\section {Beyond the Toy Model} \label{sec:beyond}

\begin{figure} 
\leftmargin=2pc
\begin{center}
\includegraphics*[width=10cm]{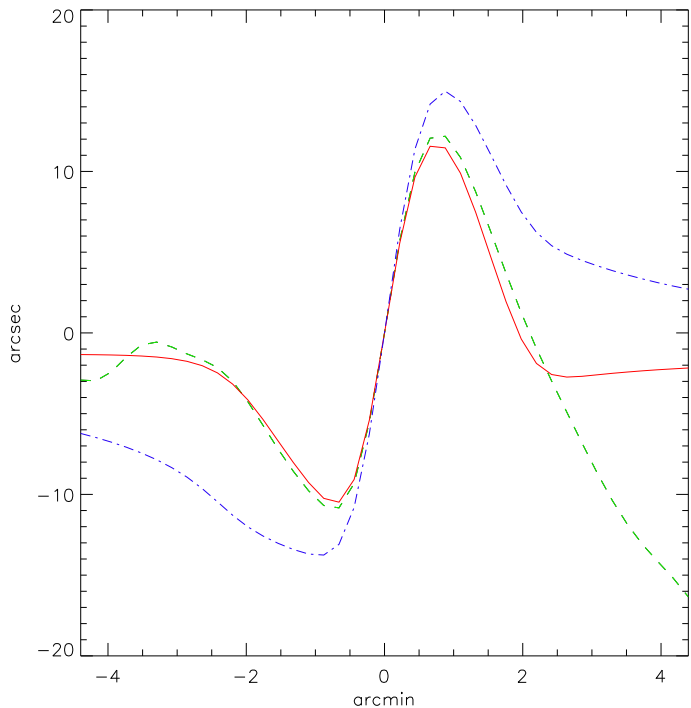}
\end{center}
\caption{The deflection angles for an isolated cluster
that occurred naturally in our simulations.  
The dot-dash line shows the deflection angle for the cluster if there were 
no other matter in the universe.
The solid line shows the lensing from the projected convergence profile at
the cluster's location on the sky.  Since lensing is dependent on the 
relative mass overdensity $\delta$, the void acts as a negative mass sheet.
Finally, the dashed line is the actual deflection angle.  
Structures away from the center of the cluster do contribute to
deflections, sometimes quite substantially.}
\label{fig:deflection}
\end{figure}

In the previous section we showed that using high resolution maps of
the CMB temperature to reconstruct stacked cluster convergence
profiles is a technique that shows some promise when examined within
the context of a simple toy model.  By employing a toy model we have
brushed aside a number of potentially important complications, perhaps
the most glaring of which is having so far ignored the fluctuations
intrinsic to the primordial CMB itself.  Although our main focus in
this section will be on stacked clusters, we nonetheless begin with
some discussion in the context of individual clusters as a natural
lead in to understanding this issue.

\begin{figure} 
\leftmargin=2pc
\begin{center}
\includegraphics*[width=6cm] {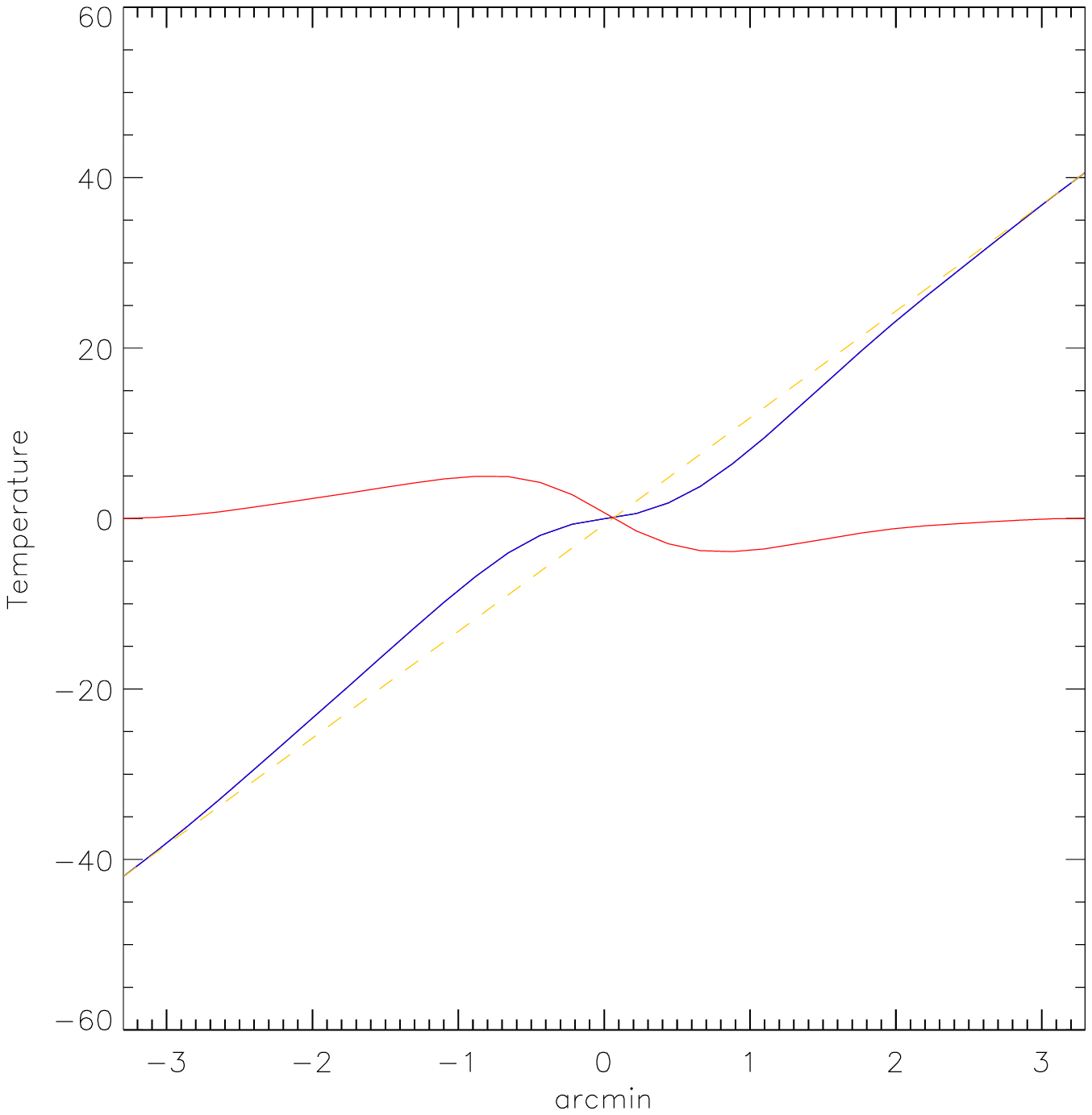}
\includegraphics*[width=6cm] {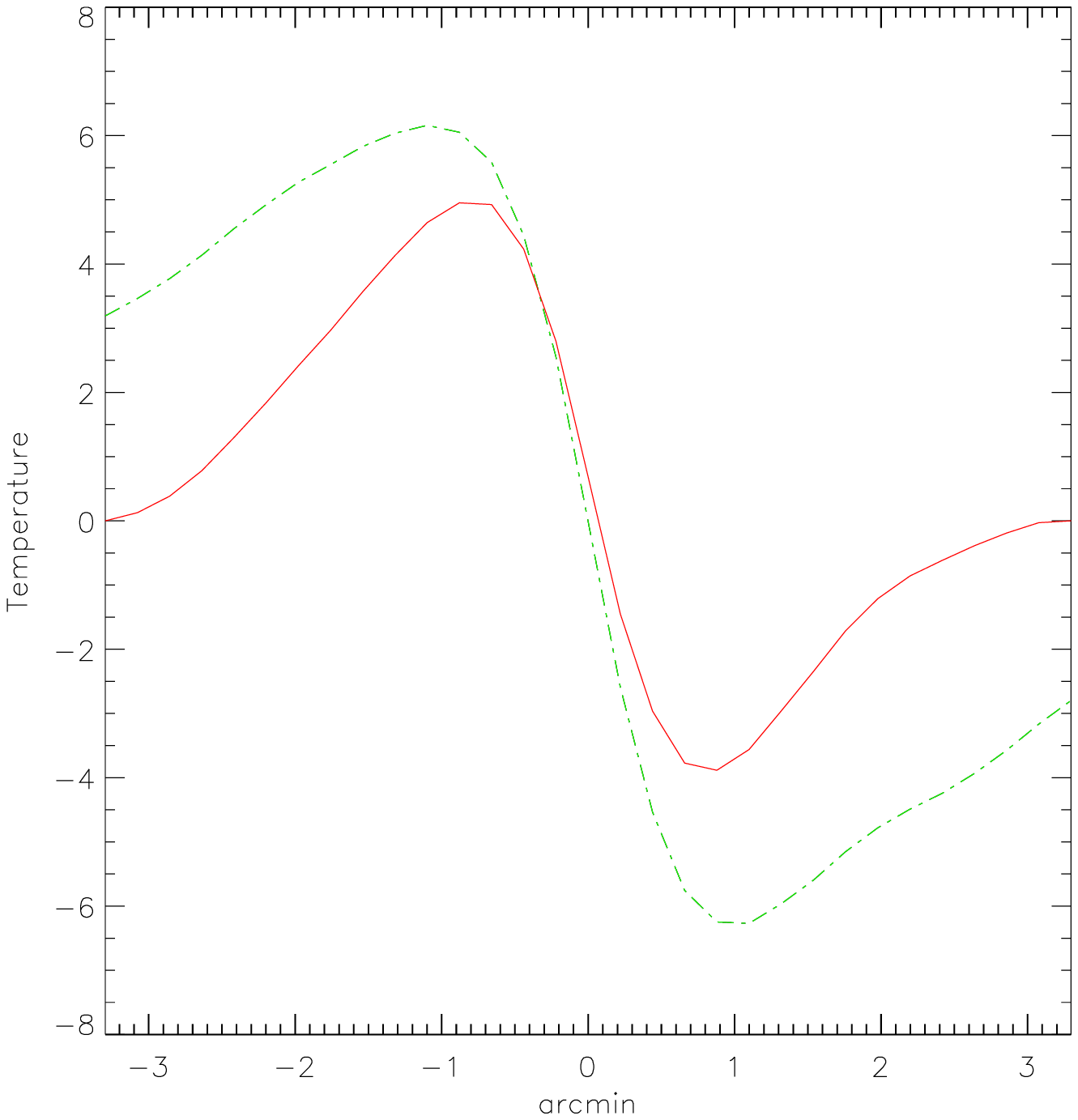}
\end{center}
\caption{The lensing signal estimated using the ansatz of
\protect\citeasnoun{SZ00}
(Left) The lensed CMB (blue) is used as an estimate of the unlensed CMB 
on either side of the cluster, and the unlensed CMB (dashed orange) is 
estimated by connecting the dots.  The difference between the two is then 
the estimate of the signal (red).
(Right) A generic feature of this method of reconstruction is its
systematic underestimate of the magnitude of the signal.  
The actual signal (dot-dash orange) is included for comparison.}
\label{fig:connectthedots}
\end{figure}

First recall the toy model ansatz of \citeasnoun{SZ00} where the CMB
temperature gradient can be approximated as constant over relevant
length scales, and the CMB can be measured far from the cluster where
the kinetic SZ and lensing by the cluster are both small.  The
unlensed CMB gradient can then be simply determined, as we illustrate
in Figure~(\ref{fig:connectthedots}), by `connecting the dots' between
the CMB on either side of the cluster.  However, the CMB cannot be
well approximated as a gradient on scales greater than a few
arcminutes even for carefully selected portions of the CMB, so to use
this technique to get a good fit for the unlensed CMB at the center of
the cluster you are forced to set the lensing signal (and therefor the
deflection angle) to zero at roughly a few arcminutes from the center
of the cluster.  Obviously the lensing effect of large clusters
extends well beyond this range, so the `connect the dots' method
systematically underestimates the cluster's mass
(Equation~\ref{eq:cluster}) as you go farther from the cluster's
center, culminating in a total cluster mass estimate equal to zero on
scales of a few arcminutes; that is, roughly at the virial radius.
This is a completely generic feature of the method, and is easily seen
in the second panel of Figure~(\ref{fig:connectthedots}) or in
Figure~(4) of \citeasnoun{HK04}.

A better method of estimating the unlensed CMB is certainly of interest.  
One obvious improvement is to extend on the original proposal of 
\citeasnoun{SZ00} by making use of something other than a simple gradient.  
For example, one could fit the CMB far from the cluster 
with a second order polynomial (but note that nearby 
structure may cause problems).  Alternatively, you could use a Wiener filter, 
as was recently suggested by \citeasnoun{HK04}.  Unfortunately, although 
the fit is somewhat improved, these methods appear to suffer from the same 
drawback as the original, and lead to estimates of the virial mass that are 
systematically low and often consistent with zero.  

A more sophisticated approach would be to iterate the fit or to include
a model for both lensing and the unlensed CMB.
This is the approach we take.
Specifically we assume that the cluster is spherically symmetric and
has an NFW profile, and we model the unlensed CMB locally as a 2nd
degree polynomial in 2D (the above mentioned nearby structure
contamination will average out when stacking).  We consider a range of
masses for the assumed NFW cluster and compute the lensing deflection
angle for each.  This is used to ``delens'' the CMB map.
The resulting map which best fits, in the $\chi^2$ sense, a second degree
polynomial (in 2D) is chosen.
A second degree polynomial has the advantage of being even about the center
of the cluster, while the lensing effect is odd, and on scales of a few
arcminutes fitting an unlensed CMB fairly well.
As a result, `bad' fits will often be glaring.

We initially tested this method on clusters `cut out' from simulations
and with no instrument noise, kinetic SZ, etc., and were able to
routinely estimate the mass to within 10\%, and to within a few
percent for our fiducial 61 cluster stack.
However, the method incurs a bias toward underestimating the mass that
becomes more significant as you include regions further from the center
of the cluster.
The bias occurs due to the failure of the 2D polynomial to accurately
represent the CMB in the region around a cluster.
One look at Figure (\ref{fig:connectthedots}) will convince the reader
of the origin of the bias.
The departure of the CMB from a constant gradient is also
odd about the center of the cluster, so that while lensing by a
cluster will degrade the ability of the polynomial to fit CMB near the
cluster, it will actually help it fit better farther away, and the two
effects compete.  Given the need for as much usable area as possible
to overcome other difficulties, it is likely that a 2D polynomial is
ultimately not the best choice for the fit.
One way to correct for this is to determine the expectation value of
the odd component in the unlensed CMB and include this in the fitting
procedure.
Alternatively with enough clusters to beat down the kSZ contamination, a
higher resolution, higher sensitivity observation would be able to work
at smaller radius where the bias is much reduced.

\section {Conclusion} \label {sec:conclusion}

In this paper we have investigated in some detail the promise of
cluster lensing of the CMB.  This idea was first introduced by
\citeasnoun{SZ00} and will soon become observationally
feasible with the imminent commissioning of the APEX-SZ telescope.  We
verify that the method of Seljak \& Zaldarriaga works well given their
assumptions, but note that these assumptions are not well satisfied in
practice.

In particular we highlight the role of the kinetic SZ signal as an important
contaminant which is spatially correlated with the cluster and spectrally
indistinguishable from the lensing signal itself.
The kSZ fluctuations, being non-Gaussian and signal-correlated, are also
an issue for reconstruction of large-scale structure as we discuss further
in \citeasnoun{AVW03}.
If the unlensed CMB can be adequately estimated, we show that the stacked
profiles and total masses are reasonably well constructed by CMB lensing,
in contrast to profiles or masses of any individual cluster.
We elucidate some of the issues in \S\ref{sec:beyond}.

Finally we make some comments about the role of polarization in lensing.
It has been emphasized before that the inclusion of polarization information
can dramatically enhance the prospects for large-scale structure
reconstruction from lensing of the CMB.  This is because lensing induces
a $B$-mode polarization signal which is otherwise absent for purely scalar,
primary fluctuations.  The large intrinsic signal, which is a source of
`noise' for lensing reconstruction, is thus absent.

It is possible that the addition of polarization information could enhance
the prospects for cluster lensing also.  To see whether the effects we have
identified are mitigated by polarization information requires a detailed
calculation.  The signal levels for polarization are much smaller, and the
spatial structure complicated as for temperature.  The kSZ effect, which is
one of our major contaminants, is also polarized.  The dominant polarization
signal comes in at order $\tau Q$ where $\tau$ is the cluster optical depth
and $Q$ the local CMB intensity quadrupole at the location of the cluster.
The resulting field is a mix of $E$- and $B$-mode signals with the
polarization pointing in the direction of the quadrupole cold lobe.
Depending on the variation of the quadrupole with distance and position in
the field, and on the cluster properties the polarization signal can be
somewhat complex.  Near each cluster though it might be possible to
significantly reduce the kSZ contamination through modeling.
We leave a detailed investigation of this question to future work.

\clearpage

\noindent {\bf Acknowledgements:}\newline
C.V. would like to thank the organizers of the workshop ``Cosmology with
Sunyaev-Zel'dovich cluster surveys'' held in Chicago in September 2003 for
allowing him the chance to present some of this work.
Additionally we would like to thank T. Chang, J. Cohn, D. Holz, B. Jain,
A. Lee, G. Smoot, M. Takada and M. Zaldarriaga for helpful discussions
about these results.
The simulations used here were performed on the IBM-SP2 at the National
Energy Research Scientific Computing Center.
This research was supported by the NSF and NASA.

\appendix
\section{The simulated map}

We construct maps of lensing convergence and the thermal and
kinetic SZ effect making use of a large, high-resolution N-body
simulation of the $\Lambda$CDM cosmology
(specifically Model 1 of \citeasnoun{YanWhiCoi}).
In this appendix we give some details of how this was done.

\subsection{The N-body simulation}

To construct the maps we need some information on the spatial distribution
and evolution of the mass in our model.  We obtain this from an N-body
simulation.
The simulation modeled a large volume of the universe, a periodic
cube $300\,h^{-1}$Mpc on a side, to ensure a good sampling of the
clusters of interest to us.  We considered only the dark matter
component which was modeled using $512^3$ particles of mass
$1.7\times 10^{10} h^{-1} M_{\odot}$.  For computation of the thermal
and kinetic SZ effects we assume that any baryonic component would
trace the dark matter, a reasonable approximation on the scales of
interest to us.

The simulation was started at $z=60$ and evolved to the present using
the TreePM code described in \citeasnoun{CodePaper}, with the full phase
space distribution dumped every $100h^{-1}$Mpc between redshifts $2>z>0$.
It is this range of redshifts which dominates the lensing and SZE signal on
the angular scales of interest to us.
The gravitational softening used is of a spline form, with a
``Plummer-equivalent'' (comoving) softening length of $20 h^{-1}$kpc.
All of the relevant cluster-scale halos contain several thousand particles to
begin to resolve sub-structure.
The simulation was performed on 128 processors of the IBM-SP2 at NERSC,
took nearly 4000 time steps and approximately 100 wall clock hours
to complete.

The maps were made in essentially the same manner as in
\citeasnoun{SchWhi} to which we refer the reader for
discussion of the various approximations.
The past lightcone was constructed by stacking the intermediate stages
of the simulation between redshifts $2>z>0$.
In order to avoid multiply sampling the same large scale structures, each
$300 h^{-1}$Mpc box has been randomly re-oriented in one of the six possible
orientations, and has furthermore been shifted by a random amount,
perpendicular to the line-of-sight, making use of the periodic boundary
conditions.
There are three time dumps per box length.
Each $300 h^{-1}$Mpc volume in the stack is made up of three segments, each
segment evolved to a later epoch than the previous one by the time it takes
light to travel $100h^{-1}$Mpc.  We chose $100 h^{-1}$Mpc as the
sampling interval because it is large enough that edge effects are minimal,
yet fine enough that the line of sight integrals are well approximated by
sums of the (static) outputs.  
Because of the periodicity, we are free to choose any of the thirds as the
oldest, cyclically permuting the other two.
This approach preserves the continuity of large-scale structure over distances
of $300 h^{-1}$Mpc without compromising the resolution in time evolution.   

We produced maps of the lensing and SZ effects.  Each map was $7.5^\circ$
on a side, with $2048^2$ square pixels each $0.22'$ on a side.
The same particle distribution and random seeds were used to construct each
of the maps, so that they represent the same patch of sky in each quantity.
We now discuss each map in turn.

\subsection{Convergence ($\kappa$) maps}

The effect of lensing was computed from maps of the convergence, $\kappa$,
assuming the weak lensing approximation (see \S\ref{sec:theory}).
This is valid except near the very center of the cluster, and so adequate
for our purposes.
The convergence was computed from the density contrast along the past
lightcone as
\begin{equation}
  \kappa = {3\over 2}\Omega_{\rm mat}H_0^2 \int d\chi\ g(\chi){\delta\over a}
\end{equation}
where $g(\chi)=\chi(\chi_s-\chi)/\chi_s$ is the lensing kernel, $\chi$ is
the comoving distance and $\chi_s\simeq 9\,h^{-1}$Gpc is the comoving
distance to the last scattering surface.
We have assumed the universe is spatially flat.
The (projected) density contrast within each $100\,h^{-1}$Mpc slice was
computed from the dark matter distribution using a spline kernel
interpolation with a smoothing length equal to the force softening in
the simulation.

\subsection{Compton-$y$ and $b$ maps}

Because the simulation contains no gas we use a semi-analytic model to include
the gas physics.
First we assume that the gas closely traces the dark matter.
This is likely a good approximation in all regions except the innermost
$O(100)$kpc of the cluster, which for clusters at cosmological
distances will be unresolved by the experiments of interest.
(e.g.~$100$kpc subtends only $0.26'$ at $z=0.5$.)
We ignore the presence of cold gas and stars in the ICM, assuming that the
mass in hot gas is $\Omega_{\rm b}/\Omega_{\rm m}$ of the total.
Second, each cluster is assumed isothermal.
We assign to each particle in a group a temperature proportional to its
mass to the $2/3$ power.

We generate Compton-$y$ maps by integrating for each pixel
\begin{equation}
  y=\int \sigma_{\rm T} n_{e}{k_{\rm B}T_{e} \over m_{e} c^2} dl
  \qquad .
\end{equation}
Here $\sigma_{\rm T}$ is the Thompson scattering cross section, and $n_e$,
$m_e$ and $T_e$ are the electron number density, mass and temperature
respectively.  We assume that within the clusters the gas is fully ionized.
The contribution from each particle is distributed over the pixels with a
spline weighting and a size equal to the smoothing length of the
simulation as for the $\kappa$ maps above.
The temperature fluctuation at frequency $\nu$ is then obtained from the
$y$-maps by
\begin{eqnarray}
{\Delta T\over T} &=&
  \phantom{-2}y \left( x{{\rm e}^x+1\over {\rm e}^x-1}-4 \right) \\
  &\simeq& -2y\qquad \mbox{for }\ x\ll 1\, ,
\end{eqnarray}
where $x=h\nu/k_BT_{CMB}\simeq\nu/56.84$GHz is the dimensionless frequency
and the second expression is valid in the Rayleigh-Jeans limit.  In what
follows we shall assume the low-frequency limit unless otherwise stated.

The Compton-$b$ maps are produced in an almost identical manner, replacing
$kT/m_ec^2$ with $v/c$, where $v$ is the line-of-sight velocity.  The
spectrum is an undistorted black body, so the temperature perturbation
is simply $\Delta T/T=-b$.

\subsection{Primary CMB anisotropies}

When needed we generate primary CMB anisotropies as a random realization
of a Gaussian field with power spectrum computed from CMBfast \cite{CMBfast}.
We generate random phases in momentum space, and assign amplitudes to each
of the $k$-modes using a distribution whose average value is the amplitude
in the CMB power spectrum.
We have used the flat sky approximation, in which the $k$-mode in momentum
space corresponds to $\ell$ value in the CMB power spectrum.

\end{document}